\documentclass[paper,11pt]{JHEP}
\usepackage[centertags]{amsmath}
\usepackage{amsfonts} \usepackage{amssymb} \usepackage{amsthm}
\usepackage{graphicx}
\usepackage{psfrag}

\def\one{{\rm 1\kern -.9mm l}}                             %

\def\beq{\begin{equation}}
\def\eeq{\end{equation}}
\def\beq{\begin{equation}}
\def\eeq{\end{equation}}
\def\beqa{\begin{eqnarray}}
\def\eeqa{\end{eqnarray}}
\newcommand{\eqa}{\begin{eqnarray}}
\newcommand{\ena}{\end{eqnarray}}
\newcommand{\eq}[1]{eq. (\ref{#1})}
\newcommand{\Tr}{\mathrm{Tr}\,}

\def\ii{\mathrm{i}}
\def\ee{\mathrm{e}}

\newcommand{\atopnew}[2]{\genfrac{}{}{0pt}{3}{#1}{#2}}
\newcommand{\cale}{\mathcal{E}}
\title{The partition function of interfaces from the Nambu-Goto
effective string theory%
\thanks{Work partially supported by the European Community's Human Potential
Programme under contract MRTN-CT-2004-005104 ``{Constituents, Fundamental
Forces and Symmetries of the Universe}'' and by the European Commission TMR 
programme HPRN-CT-2002-00325 (EUCLID)
and by the Italian M.I.U.R under contract
PRIN-2005023102 {``Strings, D-branes and Gauge Theories''}.
}}
\author{M. Bill\'o, M. Caselle, L. Ferro
\\
Dipartimento di Fisica Teorica, Universit\`a di Torino\\
and Istituto Nazionale di Fisica Nucleare - sezione di Torino \\
Via P. Giuria 1, I-10125 Torino, Italy
}
\abstract{We consider the Nambu-Goto bosonic string model as a description of the
physics of interfaces. By using the standard covariant quantization of the bosonic string,
we derive an exact expression for the partition function in dependence of the geometry of the
interface. Our expression, obtained by operatorial methods, resums the loop
expansion of the NG model in the ``physical gauge'' computed perturbatively by  
functional integral methods in the literature. Recently, very accurate Monte Carlo data
for the interface free energy in the 3d Ising model became avaliable. Our proposed expression 
compares very well to the data for values of the area sufficiently large in terms of the 
inverse string tension. This pattern is expected on theoretical grounds and agrees with 
previous analyses of other observables in the 
Ising model. 
}
\keywords{Bosonic Strings, Lattice Gauge Field Theories, Interfaces}
\preprint{DFTT/02/2006}

\begin{document}

\section{Introduction}
\label{sec:intro}
The idea that string theory may provide the effective description of confining gauge theories
in their strong-coupling regime is an old and well motivated one \cite{early,lsw,l81}; in this context, 
the string degrees of freedom describe the fluctuations of the colour flux tube. 

In the last years much effort has been devoted to test this conjecture. 
In particular, several results have been obtained in an ``effective string'' approach, 
in which the conformal anomaly due to the fact that the theory is quantized in 
a non-critical value $d\not= 26$ of the space-time dimensionality is neglected. 
In principle, this is a very problematic simplification, since conformal
invariance is at the very heart of the quantization procedure (see the
Conclusions section for some remarks on this issue). 
However, it was observed very early \cite{Olesen:1985pv} that the coefficient 
of the conformal anomaly vanishes for large distances, i.e., for world-sheets of large size in target space.
In fact, in recent years, thanks to  various improvements in lattice simulations
~\cite{lw01,cfghp97,fep00,chp03} the effective string picture has been tested with a 
very high precision and confidence \cite{chp03}-\cite{mt04} by considering observables such as Wilson 
loops and Polyakov loop correlators. It turnes out that at large inter-quark
distances and low temperatures the effective string, and in particular the simplest model, the Nambu-Goto one, correctly describes the Monte Carlo data. 
As distances are  decreased,  clear deviations from this picture are
observed~\cite{chp04,chp05,jkm03,jkm04}. 

Within the effective string framework, one can fix a ``physical gauge'' (see section \ref{subsec:ng} for some more details) and re-express a generic effective string model as a 2d (interacting) conformal field 
theory of $d-2$ bosons. In this set-up, the inverse of the product of the string tension $\sigma$ times the 
minimal area $\mathcal{A}$ of the world sheet spanned by the string represents the parameter of a loop expansion around the classical solution  for the inter-quark potential. The first term of this expansion yields the well known L\"uscher correction \cite{lsw}. 
The second term was evaluated more than twenty years ago in~\cite{Dietz:1982uc}, 
with a remarkable theoretical effort, for several different classes of effective string actions. 
Higher order corrections would require very complicated calculations and have never appeared in the literature.

Recently, a set of simulations both in SU(2) and SU(3) lattice gauge
theories (LGT's) \cite{lw02,lw04,cpr04,jkm03,jkm04,ns01,lt01,m02,maj04} 
and in the 3d gauge Ising 
model \cite{cfghp97,chp03,chp04,chp05,Caselle:2005vq}, 
thanks to powerful new algorithms, estimated the inter-quark potential with
precision high enough to distinguish among different effective string actions and to observe
the contribution of higher string modes. To compare the effective string predictions with these new
data in a meaningful way it is mandatory to go beyond the perturbative expansion. 

So far this has been done only for the simplest effective string action, the Nambu-Goto one
\cite{nambu-goto}, 
and for the cylindric geometry, which physically corresponds to the expectation value of the correlator of
two Polyakov loops. In this case it has been possible to build the partition function 
corresponding to the spectrum of the  Nambu-Goto string with the appropriate boundary conditions 
derived long ago in \cite{alvarez81,Arvis:1983fp}, and to make a successfull comparison 
with the simulations \cite{chp05,Caselle:2005vq}. In~\cite{Billo:2005iv} this partition function was 
re-derived via standard covariant quantization, showing that it indeed represents the 
exact operatorial result which re-sums the loop expansion of the model in the physical gauge. 

As a further step in this direction we derive in the present paper the exact partition 
function of the Nambu-Goto effective string for a toroidal world-sheet geometry, and 
we compare this prediction  to a set of recently published \cite{nuovomc} high precision Monte Carlo
results for the corresponding observable in the 3d gauge Ising model, namely the interface expectation 
value. 

Indeed, the toroidal geometry corresponds  in LGT's to the ``maximal 't~Hooft loop''  (see, 
for instance,~\cite{deForcrand:2005pb,Bursa:2005yv}). However a much simpler, yet physically very interesting,
observable can be associated to the same geometry if we consider a three-dimensional LGT with a discrete
abelian gauge group (like the 3d gauge Ising model). In this case the gauge model is mapped by duality into a 
three-dimensional spin model (like the 3d spin Ising model). 
In particular, the confining regime of the gauge model is
mapped into the broken symmetry phase of the spin model. Any extended gauge observable, 
like the Wilson loop or the correlator of two Polyakov loops, is mapped into a set of suitably chosen
anti-ferromagnetic bonds. The torus geometry we are interested in corresponds to the case in which the set of 
anti-ferromagnetic bonds pierces a complete slice of the lattice, i.e., to the case in which we simply impose
anti-periodic boundary conditions in one of the lattice directions. 
In a spin model with discrete symmetry group this type of boundary conditions is known to create, in the broken symmetry phase, an interface between two
different vacua. The (logarithm of the) expectation value of the slice of 
anti-ferromagnetic bonds, which we expect to be described by the Nambu-Goto string, is thus proportional to the interface free energy, an observable which has been the subject of several numerical and experimental studies in condensed matter literature. We shall briefly recall these results in section \ref{subsec:interfaces}.

To compute the partition function, we
follow the philosophy of \cite{Billo:2005iv} and resort to standard covariant quantization in the 
first order formulation of the Nambu-Goto theory, of which we briefly recall some aspects in section 
\ref{subsec:ng}. In section \ref{sec:bos_string} we integrate appropriate sectors of the bosonic string partition function over the world-sheet modular parameter. In this way we describe in an exact manner the string fluctuations around a specified target space surface, in our case representing an interface in a compact space. 
In section \ref{sec:comp_func} we show that our expression reproduces the result obtained in
\cite{Dietz:1982uc} via a perturbative expansion (up to two loops) of the NG functional integral; in fact, our expression re-sums the loop expansion.  

In section \ref{sec:mc} we compare our predictions with the data for the free energy of interfaces in the 3d Ising presented in \cite{nuovomc}. Our NG prediction agrees remarkably to the data 
for values of the area larger than (approximately) four times the inverse string tension, which is the same distance scale below which deviations from the NG model emerged in the studies of other observables cited above \cite{chp05}. 
In the range where there is agreement, we find that 
the NG result is largely dominated by the lowest level mode of the bosonic string; the free energy 
associated to this mode already accounts very well for the MC data, apart from a shift of the overall normalization. 
Though corresponding to a  single particle mode, this contribution is essentially stringy, 
since this mode pertains to a wrapped string.

\subsection{Interfaces and effective strings}
\label{subsec:interfaces}
The properties of interfaces in three-dimensional statistical systems have been a 
long-standing subject of research. In particular the interest of people 
working in the subject has been attracted by the so called ``fluid'' interfaces
whose dynamics is dominated by massless excitations
(for a review see for instance \cite{GFP,p92}). For 
this class of interfaces, thanks to the presence of long range massless modes,
microscopic details such as the  lattice structure of the spin model or the
chemical composition of the components of the binary mixture 
become irrelevant and the physics  can be rather accurately 
described by field theoretic methods. 

An effective model widely used to describe a
rough interface is the {\em  capillary wave model} (CWM) \cite{BLS,rw82}. 
Actually this model (which was proposed well before the Nambu-Goto papers)
exactly coincides~\cite{p92} with the Nambu-Goto one, since 
it assumes an effective Hamiltonian
proportional to the variation of the area of the surface with respect to the
classical solution. 

A simple realization of fluid interfaces is represented by 3d spin models.
In the broken-symmetry phase at low temperature, these models admit different 
vacua which, for a suitable choice of the boundary
conditions, can  occupy macroscopic regions and are 
separated by domain walls which behave as interfaces. For
temperatures  between the roughening and the critical one, 
interfaces are dominated by long wavelength fluctuations (i.e.\ they exactly  behave as
{\em  fluid} interfaces); all the simulations which we shall discuss below were performed in this
 region.

In these last years the 3d Ising model has played a prominent role
among the various realizations of fluid 
interfaces, for several reasons. The universality class of the Ising model
includes many physical systems, ranging from binary 
mixtures to amphiphilic membranes. Its universality class
is also the same of the $\phi^4$ theory; this allows a QFT approach to the description of 
the interface physics~\cite{m90,pv95,hm98}.
Last but not least, the Ising model, due to 
its intrinsic simplicity, allows fast and high statistics Monte Carlo
simulations, so that very precise 
comparisons can be made between theoretical predictions and numerical results.

Following this line, during the past years some high precision tests 
of the capillary wave model were performed~\cite{cfhgpv,hp97,nuovomc}. 
A remarkable agreement was found between
the numerical results for interfaces of large enough size
and the next to leading order approximation of the CWM. The present paper
represents a further step in this direction, since  we
shall be able to compare the numerical data of ~\cite{nuovomc} with the
exact prediction of the Nambu-Goto effective model for the interface free energy.

\subsection{The Nambu-Goto model}
\label{subsec:ng}
Perhaps the most natural model to describe fluctuating surfaces is the Nambu-Goto bosonic string
\cite{nambu-goto}, in which the action is proportional, via the string tension $\sigma$, to the
induced area of a surface embedded in a $d$-dimensional target space:
\begin{equation}
\label{ngaction}
S = \sigma \int d^2\xi \sqrt{\det g}~,
\hskip 0.6cm g_{\alpha\beta} = \frac{\partial X^i}{\partial\xi^\alpha}\frac{\partial X^j}{\partial \xi^\beta} G_{ij}~.
\end{equation}
Here we parametrize the surface by proper coordinates $\xi^\alpha$, and
$X^i(\xi)$ ($i=1,\ldots,d$) describes the target space position of a point specified by $\xi$.
For us the target space metric $G_{ij}$ will always be the flat one. 

The invariance under re-parametrizations of the action
(\ref{ngaction}) can be used to fix a so-called ``static'' gauge where the proper coordinates are identified with two of the target space coordinates, say $X^0$ and $X^1$. The quantum version of the NG theory 
can then be defined through the functional integration over the $d-2$ transverse d.o.f. $\vec X(X^0,X^1)$ of the gauge-fixed action. The partition function for a surface $\Sigma$ with prescribed boundary conditions
is given by
\begin{equation}
\label{ngpart}
\begin{aligned}
Z_{\Sigma} & = \int_{(\partial\Sigma)}\!\!\!\! DX^i\, \exp\left\{-\sigma \int_\Sigma dX^0 dX^1 \left(1 + (\partial_0\vec X)^2
 + (\partial_1\vec X)^2 + (\partial_0 \vec X \wedge \partial_1 \vec X)^2\right)^{\frac 12}\right\}
\\
& = \int_{(\partial\Sigma)}\!\!\!\! DX^i\,  \exp\left\{-\sigma \int_\Sigma dX^0 dX^1 \left[
1 +\frac 12 (\partial_0\vec X)^2 + \frac 12 (\partial_1\vec X)^2 + \mbox{interactions} \right]\right\}~.
\end{aligned}
\end{equation}

Expanding the square root as in the second line above, the classical area law $\exp(-\sigma\mathcal{A})$ 
(where $\mathcal{A}$ is the area of the minimal surface $\Sigma$) is singled out. It multiplies
the quantum fluctuations of the 
fields $\vec X$, which have a series of higher order (derivative) interactions. 
The functional integration can be performed perturbatively, the loop expansion parameter being $1/(\sigma\mathcal{A})$, and it 
depends on the boundary conditions imposed on the fields $\vec X$, i.e., on the topology of the boundary $\partial\Sigma$ and hence of $\Sigma$. The cases in which $\Sigma$ is a disk, a cylinder or a torus are
the ones relevant for an effective string description of, respectively, Wilson loops, Polyakov loop correlators and interfaces in a compact target space. 

The computation was carried out up to two loops in \cite{Dietz:1982uc}, see also \cite{cfhgpv,cp96}. 
The surface $\Sigma$ is taken to be a rectangle, with the opposite sides in none, one or 
both directions being identified to get the disk, the cylinder or the torus topology.
At each loop order, the result depends in a very non-trivial way on the geometry of $\Sigma$,
namely on its area $\mathcal{A}$ and on the ratio $u$ of its two sides: it typically involves non-trivial modular forms of the latter. 
The two loop result for the case of interfaces is reported here in section \ref{sec:comp_func}.

An alternative treatment of the NG model, which is the standard one described in most textbooks on string theory, see for instance~\cite{polbook},
takes advantage of the first order formulation, in which the action is simply 
\beq
\label{sac}
S = \sigma\int d\xi^0 \int_0^{2\pi} d\xi^1\,
h^{\alpha\beta}\partial_\alpha X^i \partial_\beta X^i~.
\eeq
Here $h_{\alpha\beta}$ is an independent world-sheet metric, $\xi^1\in [0,2\pi]$ parametrizes the spatial extension of the string and $\xi^0$ its proper time evolution. 
Integrating out $h$, we retrieve the NG action \eq{ngaction}. 

As it is well known, to include the process of splitting and joining of strings, i.e., to include string interactions,
one must consider world-sheets of different topology, i.e., Riemann surfaces of different genus $g$. The genus of the world-sheet represents thus the loop order in a ``string loop" expansion. 
For each fixed topology of the world-sheet, instead of integrating out $h$,
we can use re-parametrization and  Weyl invariance to put the metric in a reference form 
$\ee^\phi \hat h_{\alpha\beta}$ (conformal gauge fixing). For instance, on the sphere, i.e. at genus $g=0$, we can choose $\hat h_{\alpha\beta} = \eta_{\alpha\beta}$, while on the torus, at genus $g=1$, $\hat h_{\alpha\beta}$ is constant, but still depends on a single complex parameter $\tau$, the modulus of the torus; see later for some more details.
The scale factor $\ee^\phi$ decouples at the classical
level%
\footnote{Actually, this property persists at the quantum level only if the anomaly
parametrized by the total central charge $c= d - 26$ vanishes;
Weyl invariance is otherwise broken and the mode $\phi$, as shown by Polyakov long ago \cite{Polyakov:1981rd},
has to be thought of as a field with a Liouville-type action.
However, as argued in the Introduction and in the Conclusions section, we 
can, in first instance, neglect this effect for our purposes.} 
and the action takes then the form
\begin{equation}
\label{sac2}
S = \sigma\int d\xi^0 \int_0^{2\pi} d\xi^1\,
\hat h^{\alpha\beta}\partial_\alpha X^i \partial_\beta X^i
 + S_{\mathrm{gh.}}~,
\end{equation}
where $S_{\mathrm{gh.}}$ in \eq{sac} is the action for the ghost
and anti-ghost fields (traditionally called $c$ and $b$) that arise from the
Jacobian to fix the conformal gauge; we do not really need its explicit
expression here, see \cite{gsw} or \cite{polbook} for reviews. The modes of the Virasoro constraints 
$T_{\alpha\beta}=0$, which follow from the
$h^{\alpha\beta}$ equations of motion, generate the residual conformal invariance of  the model.
The ghost system corresponds to a CFT of central charge $c_{\mathrm{gh.}} = -26$.
The fields $X^i(\tau,\sigma)$, with $i=1,\ldots,d$, describe the embedding of
the string world-sheet in the target space and form the simple, well-known two-dimensional CFT of $d$
free bosons. 

\section{The partition function for the interface from bosonic strings}
\label{sec:bos_string}
In the present section we want to describe the fluctuations of an interface in a toroidal target space 
$T^d$ by means of standard closed bosonic string theory. We use the standard first order formulation discussed in section \ref{subsec:ng} and we specify the periodicity of the target space coordinates to be $x^i\sim x^i + L^i$ ($i=1,\ldots,d$). 

The partition function for the bosonic string on the target torus $T^d$ is expressed
as
\begin{equation}
\label{bos1}
\mathcal{Z}^{(d)} = \int \frac{d^2\tau}{\tau_2} \, Z^{(d)}(q,\bar q)\, Z^{\mathrm{gh}}(q,\bar q).
\end{equation}
Here $\tau = \tau_1 + \ii\tau_2$ is the modular parameter of the world-sheet, 
which is a surface of genus $g=1$, i.e., a torus. Moreover, $Z^{(d)}(q,\bar q)$ is the
CFT partition function of the $d$ compact bosons $X^i$ defined on such a world-sheet. 
In an operatorial formulation, this reads
\begin{equation}
\label{bos2}
Z^{(d)}(q,\bar q) = \Tr\, q^{L_0 - \frac{d}{24}}\, \bar q^{\tilde L_0 - \frac{d}{24}}~,
\end{equation}
where 
\begin{equation}
\label{bos2bis}
q = \exp (2\pi\ii\tau)~,
\hskip 0.4cm
\bar q = \exp (-2\pi\ii\bar\tau)~.
\end{equation}
$L_0$ and $\tilde L_0$ (particular modes of the Virasoro constraints) are the left
and right-moving dilation generators. With 
$Z^{\mathrm{gh}}(q,\bar q)$ we denote the CFT partition function for the ghost system, defined on the same world-sheet.

The modular parameter $\tau$ is the Teichmuller parameter of the world-sheet surface: as discussed
in subsection \ref{subsec:ng}, using Weyl invariance and diffeomorphisms we can choose the reference metric $\hat h_{\alpha\beta}$ to be constant and of unit determinant, 
but the complex parameter $\tau$, with $\mathrm{Im}\tau\ge 0$,  which characterizes the complex structure, i.e., the shape of the torus, cannot be fixed.
It is thus necessary to integrate over it, as indicated in \eq{bos1}, the integration domain%
\footnote{We will discuss later the issue of discrete modular transformations acting on $\tau$.} 
being the upper half-plane. In the Polyakov approach \cite{Polyakov:1981rd},
this integration is the remnant of the functional integration over the independent world-sheet metric $h_{\alpha\beta}$ after the invariances of the model have been used as described above.
The measure used in \eq{bos3} ensures, as we will see, the modular invariance of the integrand. 

The CFT partition function for a single boson defined on a circle 
\begin{equation}
\label{bos3} X(\xi^0,\xi^1) \sim X(\xi^0,\xi^1) + L
\end{equation}
is given, with the action defined as in \eq{bos1}, by%
\footnote{With an abuse of notation, and for the sake of convenience, we will sometimes denote the Dedekind eta function $\eta(\tau)$ defined in 
\eq{defeta1} as $\eta(q)$, where $q=\exp(2\pi\ii\tau)$.} 
\begin{equation}
\label{bos4}
Z(q,\bar q) = \Tr\, q^{L_0 - \frac{d}{24}}\, \bar q^{\tilde L_0 - \frac{d}{24}}
= \sum_{n,w\in\mathbb{Z}} q^{\frac{1}{8\pi\sigma}\left(\frac{2\pi n}{L} + \sigma w L\right)^2}
\bar q^{\frac{1}{8\pi\sigma}\left(\frac{2\pi n}{L} - \sigma w L\right)^2}
\frac{1}{\eta(q)}\, \frac{1}{\eta(\bar q)}~.
\end{equation}
The integers $n$ and $w$ are zero-modes of the field $X$, describing respectively its discrete momentum
$p = 2\pi n/L$ and its winding around the compact target space: $X$ must be periodic in $\xi^1$, but 
the target space identification \eq{bos3} allows the possibility that
\begin{equation}
\label{bos5}
X(\xi^0,\xi^1 + 2\pi) = X(\xi^0,\xi^1) + w L~.
\end{equation}
The factors of $1/\eta(q)$ and $1/\eta(\bar q)$ result from the trace over the left and right moving 
non-zero modes, which after canonical quantization become just bosonic oscillators contributing 
to $L_0$ and $\tilde L_0$ their total occupation numbers.

The Hamiltonian trace \eq{bos4} can be re-summed \'a la Poisson, see \eq{poisson_res}, over the integer $n$, after which it becomes
\begin{equation}
\label{bos6}
Z(q,\bar q) = \sqrt{\frac{\sigma}{2\pi}}L\sum_{m,w\in\mathbb{Z}} 
\ee^{- \frac{\sigma L^2}{2\tau_2} |m - \tau w|^2}
\frac{1}{\sqrt{\tau_2}\eta(q)\eta(\bar q)}~,
\end{equation}
an expression which is naturally obtained from the path-integral formulation. The discrete
sum over $m,w$ represents the sum over ``world-sheet instantons'', namely classical solutions of the field $X$ which, beside their wrapping number $w$ over the $\xi^1$ direction, are characterized also by their wrapping $m$ along the compact propagation direction:
\begin{equation}
\label{bos7}
X(\xi^0 + 2\pi\tau_2,\xi^1 + 2\pi\tau_1) = X(\xi^0,\xi^1) + m L~.
\end{equation}

The form \eq{bos6} of $Z(q,\bar q)$ makes its modular invariance manifest. In fact, the combination
$\sqrt{\tau_2}\eta(q)\eta(\bar q)$ is modular invariant, as it follows from the properties of the Dedekind eta function 
given in \eq{etaS}. Moreover, from the exponential term we infer the effect of modular transformations of the parameter $\tau$ on the wrapping integers $w,m$: they act as $\mathrm{SL}(2,\mathbb{Z})$ matrices on the vector $(m,w)$. In particular, the $S$ and $T$ generators of the modular group are represented as follows:
\begin{align}
\label{bos8}
S:&\null & \tau &\to -\frac{1}{\tau}~, & 
\begin{pmatrix}
m \\ w
\end{pmatrix}
 &\to 
\begin{pmatrix}
0 & -1\\ 1 & 0
\end{pmatrix}
\begin{pmatrix}
m \\ w
\end{pmatrix}~,
\\
T:&\null & \tau &\to \tau+1~, & 
\begin{pmatrix}
m \\ w
\end{pmatrix}
 &\to 
\begin{pmatrix}
1 & -1\\ 0 & 1
\end{pmatrix}
\begin{pmatrix}
m \\ w
\end{pmatrix}~.
\end{align}
This allows to reabsorb the effect of modular transformations by relabelling the sums over $m$ and $w$. 

The partition function for the ghost system is given by
\begin{equation}
\label{bos9}
Z^{\mathrm{gh}}(q,\bar q) = \left(\eta(q)\eta(\bar q)\right)^2~,
\end{equation}
namely it coincides with the inverse of the non zero-mode contributions of two bosons.
Notice that the string partition function \eq{bos1} can be rewritten, substituting the
above expression for $Z^{\mathrm{gh}}(q,\bar q)$, in an explicitly modular-invariant way:
\begin{equation}
\label{bos10}
\mathcal{Z}^{(d)} = \int \frac{d^2\tau}{(\tau_2)^2} \, \left(\sqrt{\tau_2}\eta(q)\eta(\bar q)\right)^2\, 
Z^{(d)}(q,\bar q)~,
\end{equation}
so that the integration over the modular parameter $\tau$ in \eq{bos10} has to be restricted to
the fundamental cell of the modular group. 
Indeed, the Poincar\'e measure $d^2\tau/(\tau_2)^2$ 
and the combination $\sqrt{\tau_2}\eta(q)\eta(\bar q)$ are modular invariant.
The bosonic partition function $Z^{(d)}(q,\bar q)$ is also modular invariant, being 
the product of $d$ expressions of the type \eq{bos4}. It depends
on integers $n^i$ and $w^i$, the discrete momentum and winding number 
for each direction. For each direction $i$, we can Poisson re-sum over the discrete momentum $n^i$, 
as in \eq{bos6}, replacing it with the topological number $m^i$. 

\begin{figure}
\begin{center}

\begin{picture}(0,0)%
\includegraphics{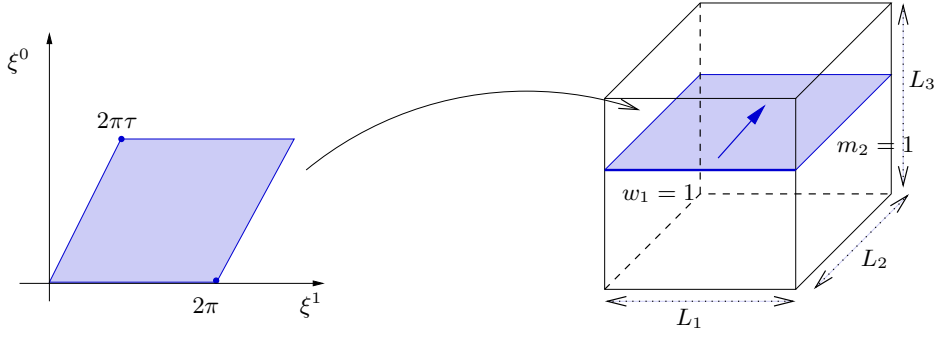}%
\end{picture}%
\setlength{\unitlength}{1973sp}%
\begingroup\makeatletter\ifx\SetFigFont\undefined%
\gdef\SetFigFont#1#2#3#4#5{%
  \reset@font\fontsize{#1}{#2pt}%
  \fontfamily{#3}\fontseries{#4}\fontshape{#5}%
  \selectfont}%
\fi\endgroup%
\begin{picture}(12048,4153)(676,-3667)
\put(12001,-586){\makebox(0,0)[lb]{\smash{{\SetFigFont{9}{10.8}{\rmdefault}{\mddefault}{\updefault}$L_3$}}}}
\put(8401,-2011){\makebox(0,0)[lb]{\smash{{\SetFigFont{9}{10.8}{\rmdefault}{\mddefault}{\updefault}$w_1=1$}}}}
\put(11101,-1411){\makebox(0,0)[lb]{\smash{{\SetFigFont{9}{10.8}{\rmdefault}{\mddefault}{\updefault}$m_2=1$}}}}
\put(9076,-3586){\makebox(0,0)[lb]{\smash{{\SetFigFont{9}{10.8}{\rmdefault}{\mddefault}{\updefault}$L_1$}}}}
\put(11401,-2836){\makebox(0,0)[lb]{\smash{{\SetFigFont{9}{10.8}{\rmdefault}{\mddefault}{\updefault}$L_2$}}}}
\put(676,-361){\makebox(0,0)[lb]{\smash{{\SetFigFont{9}{10.8}{\rmdefault}{\mddefault}{\updefault}$\xi^0$}}}}
\put(4351,-3436){\makebox(0,0)[lb]{\smash{{\SetFigFont{9}{10.8}{\rmdefault}{\mddefault}{\updefault}$\xi^1$}}}}
\put(3001,-3436){\makebox(0,0)[lb]{\smash{{\SetFigFont{9}{10.8}{\rmdefault}{\mddefault}{\updefault}$2\pi$}}}}
\put(1801,-1111){\makebox(0,0)[lb]{\smash{{\SetFigFont{9}{10.8}{\rmdefault}{\mddefault}{\updefault}$2\pi\tau$}}}}
\end{picture}%

\end{center}
\caption{\label{fig:sector}
\small The mapping of the toroidal string world-sheet, of modular parameter $\tau$, into the target space
is organized in many distinct sectors, labeled by the integers $w_i$ and $m_i$ (see the text). By 
selecting the sector with (say) $w_1=1$ and $m_2=1$ we are considering the fluctuations of an 
extended interface, which is a torus because of the target space periodicity.}
\end{figure}

\begin{figure}
\begin{center}

\begin{picture}(0,0)%
\includegraphics{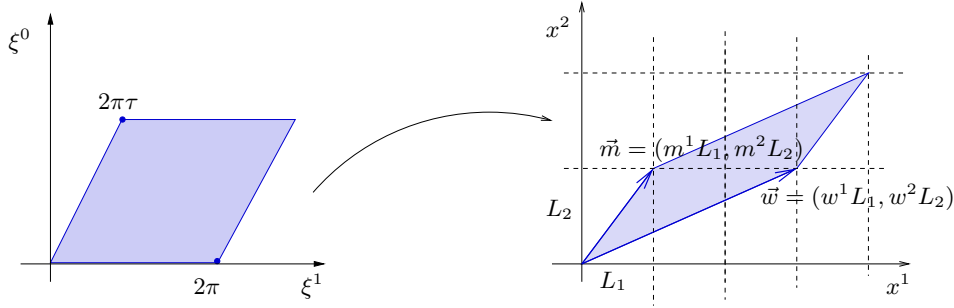}%
\end{picture}%
\setlength{\unitlength}{1973sp}%
\begingroup\makeatletter\ifx\SetFigFont\undefined%
\gdef\SetFigFont#1#2#3#4#5{%
  \reset@font\fontsize{#1}{#2pt}%
  \fontfamily{#3}\fontseries{#4}\fontshape{#5}%
  \selectfont}%
\fi\endgroup%
\begin{picture}(11717,3849)(676,-3598)
\put(676,-361){\makebox(0,0)[lb]{\smash{{\SetFigFont{9}{10.8}{\rmdefault}{\mddefault}{\updefault}$\xi^0$}}}}
\put(4351,-3436){\makebox(0,0)[lb]{\smash{{\SetFigFont{9}{10.8}{\rmdefault}{\mddefault}{\updefault}$\xi^1$}}}}
\put(3001,-3436){\makebox(0,0)[lb]{\smash{{\SetFigFont{9}{10.8}{\rmdefault}{\mddefault}{\updefault}$2\pi$}}}}
\put(1801,-1111){\makebox(0,0)[lb]{\smash{{\SetFigFont{9}{10.8}{\rmdefault}{\mddefault}{\updefault}$2\pi\tau$}}}}
\put(11701,-3436){\makebox(0,0)[lb]{\smash{{\SetFigFont{9}{10.8}{\rmdefault}{\mddefault}{\updefault}$x^1$}}}}
\put(7426,-211){\makebox(0,0)[lb]{\smash{{\SetFigFont{9}{10.8}{\rmdefault}{\mddefault}{\updefault}$x^2$}}}}
\put(8101,-3361){\makebox(0,0)[lb]{\smash{{\SetFigFont{9}{10.8}{\rmdefault}{\mddefault}{\updefault}$L_1$}}}}
\put(7426,-2461){\makebox(0,0)[lb]{\smash{{\SetFigFont{9}{10.8}{\rmdefault}{\mddefault}{\updefault}$L_2$}}}}
\put(8101,-1711){\makebox(0,0)[lb]{\smash{{\SetFigFont{9}{10.8}{\rmdefault}{\mddefault}{\updefault}$\vec m = (m^1 L_1,m^2 L_2)$}}}}
\put(10126,-2311){\makebox(0,0)[lb]{\smash{{\SetFigFont{9}{10.8}{\rmdefault}{\mddefault}{\updefault}$\vec w = (w^1 L_1,w^2 L_2)$}}}}
\end{picture}%

\caption{\label{fig:maps}
\small Consider an embedding of the world-sheet torus into the target space $T^2$ aligned along the directions $x^1,x^2$, characterized by the wrapping numbers $(m^1,w^1)$ and $(m^2,w^2)$. 
It corresponds, in the covering space of this $T^2$, to a parallelogram defined by the vectors 
$\vec w = (w^1 L_1,w^2 L_2)$ and $\vec m = (m^1 L_1,m^2 L_2)$, whose area is $\vec w\wedge \vec m = L_1 L_2(w^1 m^2 - w^2 m^1)$.}
\end{center}
\end{figure}

We want to single out the contributions to the partition function \eq{bos3}
which describe the fluctuations of an interface aligned along a  two-cycle $T^2$ inside $T^d$, 
say the one in the $x^1,x^2$ directions. The world-sheet torus parametrized by $\xi^0,\xi^1$ must be
mapped onto the target space torus by embedding functions $X^1(\xi^0,\xi^1),X^2(\xi^0,\xi^1)$ with 
non-trivial wrapping numbers $(m^1,w^1)$ and $(m^2,w^2)$. The wrapping numbers of $X^i(\xi^0,\xi^1)$, $i>2$, must instead vanish. The minimal area spanned by such a wrapped torus is given by (see Fig. \ref{fig:maps})
\begin{equation}
\label{areamw}
L_1 L_2\, 
(w^1 m^2 - m^1 w^2) =
L_1 L_2
\begin{pmatrix}
m^1 & w^1 
\end{pmatrix}
\begin{pmatrix}
0 & -1 \\
1 & 0 
\end{pmatrix}
\begin{pmatrix}
m^2 \\ w^2 
\end{pmatrix}~.
\end{equation}

\begin{figure}
\begin{center}

\begin{picture}(0,0)%
\includegraphics{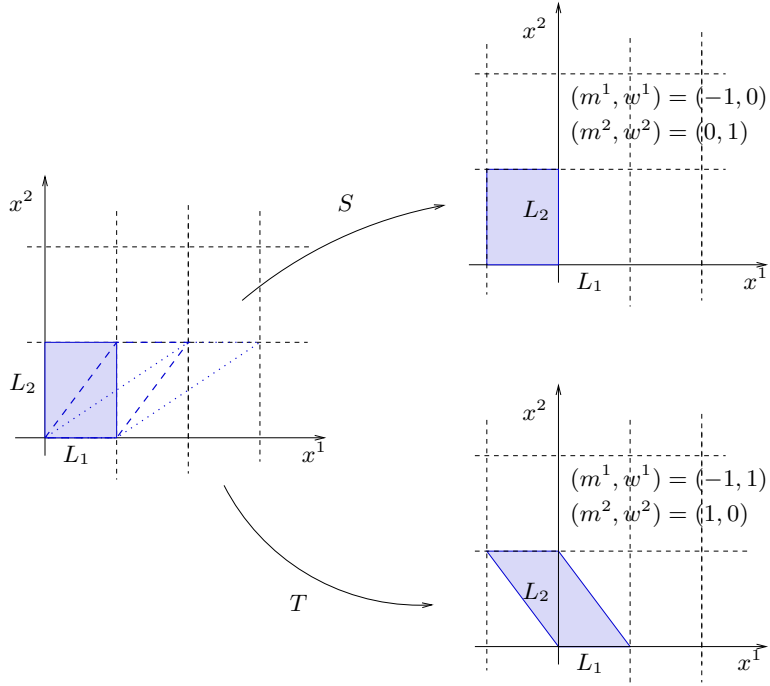}%
\end{picture}%
\setlength{\unitlength}{1973sp}%
\begingroup\makeatletter\ifx\SetFigFont\undefined%
\gdef\SetFigFont#1#2#3#4#5{%
  \reset@font\fontsize{#1}{#2pt}%
  \fontfamily{#3}\fontseries{#4}\fontshape{#5}%
  \selectfont}%
\fi\endgroup%
\begin{picture}(9917,8649)(301,-8098)
\put(7351,-736){\makebox(0,0)[lb]{\smash{{\SetFigFont{9}{10.8}{\rmdefault}{\mddefault}{\updefault}$(m^1,w^1)=(-1,0)$ }}}}
\put(7351,-1186){\makebox(0,0)[lb]{\smash{{\SetFigFont{9}{10.8}{\rmdefault}{\mddefault}{\updefault}$(m^2,w^2)=(0,1)$ }}}}
\put(301,-2086){\makebox(0,0)[lb]{\smash{{\SetFigFont{9}{10.8}{\rmdefault}{\mddefault}{\updefault}$x^2$}}}}
\put(976,-5236){\makebox(0,0)[lb]{\smash{{\SetFigFont{9}{10.8}{\rmdefault}{\mddefault}{\updefault}$L_1$}}}}
\put(301,-4336){\makebox(0,0)[lb]{\smash{{\SetFigFont{9}{10.8}{\rmdefault}{\mddefault}{\updefault}$L_2$}}}}
\put(3976,-5236){\makebox(0,0)[lb]{\smash{{\SetFigFont{9}{10.8}{\rmdefault}{\mddefault}{\updefault}$x^1$}}}}
\put(6751, 89){\makebox(0,0)[lb]{\smash{{\SetFigFont{9}{10.8}{\rmdefault}{\mddefault}{\updefault}$x^2$}}}}
\put(7426,-3061){\makebox(0,0)[lb]{\smash{{\SetFigFont{9}{10.8}{\rmdefault}{\mddefault}{\updefault}$L_1$}}}}
\put(6751,-2161){\makebox(0,0)[lb]{\smash{{\SetFigFont{9}{10.8}{\rmdefault}{\mddefault}{\updefault}$L_2$}}}}
\put(6751,-4711){\makebox(0,0)[lb]{\smash{{\SetFigFont{9}{10.8}{\rmdefault}{\mddefault}{\updefault}$x^2$}}}}
\put(7426,-7861){\makebox(0,0)[lb]{\smash{{\SetFigFont{9}{10.8}{\rmdefault}{\mddefault}{\updefault}$L_1$}}}}
\put(6751,-6961){\makebox(0,0)[lb]{\smash{{\SetFigFont{9}{10.8}{\rmdefault}{\mddefault}{\updefault}$L_2$}}}}
\put(9526,-3061){\makebox(0,0)[lb]{\smash{{\SetFigFont{9}{10.8}{\rmdefault}{\mddefault}{\updefault}$x^1$}}}}
\put(9451,-7861){\makebox(0,0)[lb]{\smash{{\SetFigFont{9}{10.8}{\rmdefault}{\mddefault}{\updefault}$x^1$}}}}
\put(4426,-2086){\makebox(0,0)[lb]{\smash{{\SetFigFont{9}{10.8}{\rmdefault}{\mddefault}{\updefault}$S$}}}}
\put(3826,-7111){\makebox(0,0)[lb]{\smash{{\SetFigFont{9}{10.8}{\rmdefault}{\mddefault}{\updefault}$T$}}}}
\put(7351,-5536){\makebox(0,0)[lb]{\smash{{\SetFigFont{9}{10.8}{\rmdefault}{\mddefault}{\updefault}$(m^1,w^1)=(-1,1)$ }}}}
\put(7351,-5986){\makebox(0,0)[lb]{\smash{{\SetFigFont{9}{10.8}{\rmdefault}{\mddefault}{\updefault}$(m^2,w^2)=(1,0)$ }}}}
\end{picture}%

\caption{\label{fig:mod_transf}
\small On the left, (some of) the coverings corresponding to the chosen sector of the partition function are depicted. Beside the one which exactly corresponds to the fundamental cell of the target space $T^2$, i.e. the case with $(m^1,w^1)=(0,1)$ and $(m^2,w^2)=(1,0)$ there are  ``slanted'' ones corresponding to generic values of $m^1$.
The generators $S$ and $T$ of the world-sheet modular group, which act by $\mathrm{SL}(2,\mathbb{Z})$ matrices as indicated in \eq{bos8}, map these coverings to different ones with the same area. For instance, on the right, we draw the $S$- and $T$-transform of the ``fundamental'' covering discussed above (the solid one in the leftmost drawing).}
\end{center}
\end{figure}

A particular sector which contributes to such a target-space configuration can be obtained 
(see Figure \ref{fig:sector})
by considering a string winding once in, say, the $x^1$ direction:
\begin{equation}
\label{bos11}
w_1 = 1~,\hskip 0.4cm
w_2 = w_3 = \ldots = w_d = 0
\end{equation}
and, upon Poisson re-summation along the directions $x^2$ to $x^d$, selecting the integers
\begin{equation}
\label{bos12}
m_2 = 1~,\hskip 0.4cm
m_3 = m_4 = \ldots = m_d = 0~.
\end{equation}
This corresponds to wrapping numbers in the $x^1,x^2$ directions given by $(m^1,w^1)=(m^1,1)$
and $(m^2,w^2) = (1,0)$ and hence, according to \eq{areamw}, to a minimal area $L_1 L_2$: such a configuration covers the two-cycle in the $x^1,x^2$ directions just once. We have not fixed the value of the wrapping number $m^1$, therefore the sector we consider contains infinite embeddings that corresponds to ``slanted'' coverings, see Fig. \ref{fig:mod_transf}.

The choice $w_i=m_i=0$ for $i=3,\ldots, d$ is preserved under the action 
\eq{bos8} of the modular group. Moreover, this $\mathrm{SL}(2,\mathbb{Z})$ action preserves
the area \eq{areamw}, as it is easy to see%
\footnote{The expression of the area \eq{areamw} contains the simplectic product of the two vectors
$(m^1,w^1)$ and $(m^2,w^2)$. The simultaneous action of $\mathrm{SL}(2,\mathbb{Z})$ on both vectors
preserves this symplectic product: $\mathrm{SL}(2)\sim \mathrm{Sp}(1)$.}. 
The orbit of the modular group which contains the configurations chosen in eq.s (\ref{bos11}-\ref{bos12}) spans thus infinite other string configurations characterized by different wrapping numbers in the directions $x^1$, $x^2$; all these configurations, however, always wrap the $T^2$ in target space just once, and correspond to equivalent descriptions of the interface; see Fig. \ref{fig:mod_transf}.

To obtain the partition function for this interface we should include all the wrapping numbers $(m^1,w^1)$ and $(m^2,w^2)$ which are in the modular orbit of the configurations in eq.s (\ref{bos11}-\ref{bos12}), but we should 
integrate the $\tau$ parameter in \eq{bos13} over the fundamental cell of the modular group only, as usual, to avoid overcounting. We can, equivalently, consider only the configurations of eq.s (\ref{bos11}-\ref{bos12}), fixing
in this way the degeneracy associated to modular transformations, and integrate $\tau$ all over the upper half-plane, which contains all the images under modular transformations of the fundamental cell.

The second choice turns out to be convenient, as the integrals become very simple.
In this way, we get the following expression for the interface partition function:
\begin{equation}
\label{bos13}
\mathcal{I}^{(d)}\!  = \!
\int\! \frac{d^2\tau}{(\tau_2)^{\frac{d+1}{2}}} 
\left[\frac{1}{\eta(q)}\frac{1}{\eta(\bar{q})}\right]^{d-2}\!
\sum_{n_1\in\mathbb{Z}} q^{\frac{1}{8\pi\sigma}\left(\frac{2\pi n_1}{L_1} +  \sigma L_1\right)^{2}}
\bar q^{\frac{1}{8\pi\sigma}\left(\frac{2\pi n_1}{L_1} -  \sigma L_1\right)^{2}}
\prod_{i=2}^d \left(\sqrt{\frac{\sigma}{2\pi}} L_i\right)
\ee^{-\frac{\sigma L_2^2}{2\tau_2}}~.
\end{equation}

Expanding in series the Dedekind's $\eta$ functions:
\begin{equation}
\label{bos14}
\left[\eta(q)\right]^{2-d} = \sum_{k=0}^{\infty}c_k q^{k-\frac{d-2}{24}}~,
\end{equation}
and expressing $q$ and $\bar q$ in terms of $\tau$, we rewrite \eq{bos13} as
\begin{equation}
\label{bos15}
\begin{aligned}
\mathcal{I}^{(d)}  & = \prod_{i=2}^d \left(\sqrt{\frac{\sigma}{2\pi}}L_i\right)
\sum_{k,k'=0}^\infty \sum_{n_1\in \mathbb{Z}} c_k c_{k'}\int_{-\infty}^\infty d\tau_1
\ee^{2\pi\ii(k-k'+n_1)} \int_0^\infty \frac{d\tau_2}{(\tau_2)^{\frac{d+1}{2}}}
\\
& \times
\exp\left\{ -\tau_{2}\left[\frac{\sigma L_{1}^{2}}{2} + \frac{2\pi^{2} n_{1}^{2}}{\sigma L_{1}^{2}} + 2\pi(k+k'-\frac{d-2}{12})\right]-\frac{1}{\tau_{2}}\left[\frac{\sigma L_{2}^{2}}{2}\right]\right\}~.
\end{aligned}
\end{equation}
The integration over $\tau_2$ can be carried out in terms of modified Bessel functions
using the formula
\begin{equation}
\label{bos17}
\int_0^\infty\frac{d\tau_{2}}{\tau_{2}^{\frac{d+1}{2}}}\exp\left\{ -A^{2}\tau_{2}-\frac{B^{2}}{\tau_{2}}\right\} =2\left(\frac{A}{B}\right)^{\frac{d-1}{2}}K_{\frac{d-1}{2}}(2AB)~,
\end{equation}
with 
\begin{equation} 
\label{bos18}
A  =\sqrt{\frac{\sigma}{2}} L_{1}\, \cale~,
\hskip 0.4cm
B  = \sqrt{\frac{\sigma}{2}} L_{2}
\end{equation}
and
\begin{equation}
\label{bos19}
\cale  =
\sqrt{1 + \frac{4\pi}{\sigma L_1^2}(k+k'-\frac{d-2}{12}) + 
\frac{4\pi^{2}\, n_{1}^{2}}{\sigma^2 L_{1}^{4}}}\\
= \sqrt{1 + \frac{4\pi\, u}{\sigma \mathcal{A}}(k+k'-\frac{d-2}{12}) + 
\frac{4\pi^{2} \, u^2\, n_{1}^{2}}{(\sigma\mathcal{A})^2}}~.
\end{equation}
In the second step, we introduced the area $\mathcal{A}$ and the modular parameter $\ii u$ of the torus swept out by the string in the target space, namely the interface:
\begin{equation}
\label{Au}
\mathcal{A} = L_1 L_2~, \hskip 0.4cm
u = \frac{L_2}{L_1}~.
\end{equation}

The integration over $\tau_1$ produces a $\delta(k -k' + n_1)$ factor, and thus,
using Eq.s \ref{bos17}--\ref{bos18}, we rewrite the interface partition function 
\eq{bos15} as
\begin{equation}
\label{boskkp}
\mathcal{I}^{(d)} = 2 \left(\frac{\sigma}{2\pi}\right)^ {\frac{d-2}{2}}\, V_T \, \sqrt{\sigma\mathcal{A}u}
\sum_{k,k'=0}^\infty  c_k c_{k'} 
\left(\frac{\cale}{u}\right)^{\frac{d-1}{2}}\, K_{\frac{d-1}{2}}\left(\sigma\mathcal{A}\cale\right)~,
\end{equation}
where $\cale$, introduced in \eq{bos19}, is now to be written replacing $k - k'$ for $n_1$.
In \eq{boskkp} we have introduced the transverse volume 
$V_T = \prod_{i=3}^d L_i$, and we have re-written the prefactor $\sqrt{\sigma}L_2$ as 
$\sqrt{\sigma\mathcal{A}u}$.

Eq. (\ref{boskkp})
gives the exact expression for the fluctuations of the interface aligned along 
the $x^1,x^2$ directions, in our hypothesis that they be described from standard bosonic string theory.
This expression depends only on the target space geometric data, namely the transverse volume $V_T$ and
the area $\mathcal{A}$ and the shape parameter $u$ of the interface.

Considering the leading exponential behaviour of the Bessel functions only, \eq{boskkp} would reduce
to a partition function constructed by summing over closed string states, characterized by the left- and right-moving occupation numbers $k$ and $k'$ and by the momentum $n_1$ which is fixed to $k-k'$ by the level-matching condition, with the usual bosonic multiplicities $c_k$, $c_{k'}$ and with exponential weights
$\exp(-L_2 E_{k,k'})$, where the energies are given by
\begin{equation}
\label{spectrum}
E_{k,k'} = \sigma L_1 \cale = \sqrt{\sigma^2 L_1^2 + 4\pi\sigma (k+k'-\frac{d-2}{12}) + 
\frac{4\pi^{2}\, n_{1}^{2}}{L_{1}^{2}}}~.
\end{equation}
This spectrum substantially agrees with the expression proposed in \cite{Kuti:2005xg}; however, the correct expression of the partition function, \eq{boskkp}, involves Bessel functions rather than exponentials 
and it also contains extra factors of $\cale$. These modifications are crucial in reproducing correctly the loop expansion of the functional approach, as we shall see in the next section.

\section{Comparison with the functional integral approach} 
\label{sec:comp_func}
As discussed in the Introduction, the Nambu-Goto interface partition function, whose expression \eq{bos21}
we derived by operatorial methods in the first-order formalism, can also be computed in a functional integral 
approach with a physical gauge-fixing. This leaves just the $d-2$ bosonic d.o.f. corresponding to the transverse fluctuations of the interface, but the action is not a free one. One can 
evaluate the path integral perturbatively, the loop expansion parameter being the inverse of $\sigma\mathcal{A}$ \cite{Dietz:1982uc}. The result of this computation
up to the 2-d loop order was given in
\cite{Dietz:1982uc} and reads
\begin{equation}
\label{df1}
\mathcal{I}^{(d)} \propto\sigma^{\frac{d-2}{2}} \frac{\ee^{-\sigma\mathcal{A}}}{\left[\sqrt{u}
\eta^2(\ii u)\right]^{d-2}} \left\{1 + \frac{f_1(u)}{\sigma\mathcal{A}} + \ldots\right\}~. 
\end{equation}
where the dots stand for higher loop contributions and
\begin{equation}
\label{fuis}
f_1(u) = \frac{(d-2)^2}{2} \left[\left(\frac{\pi}{6}\right)^2 u^2 E_2^2(\ii u) - \frac{\pi}{6} E_2(\ii u)\right]
+ \frac{d(d-2)}{8}~.
\end{equation}
Here $E_2$ denotes the 2nd Eisenstein series (see Appendix \ref{app:useful}). 
Actually, the constant term in \eq{fuis} is different from the one given in 
\cite{Dietz:1982uc}; in Appendix \ref{app:2loop} we show, by reconsidering the computation of the second loop 
term, that in fact the correct result is the one we quote in \eq{fuis}%
\footnote{It is interesting to observe that the missing contribution in \cite{Dietz:1982uc} is proportional to $(d-3)$ and thus disappears in three dimensions. 
This is the reason for which it was not found in the calculations reported
in~\cite{cfhgpv,cp96} which evaluated this 2 loop correction with three other types of regularization in the $d=3$ case. Moreover the difference is modular invariant and thus it could not be detected by Dietz and Filk in the tests of their calculation which they made in \cite{Dietz:1982uc}.}

Our exact expression \eq{boskkp} should reproduce the perturbative expansion of the functional integral result when asymptotically expanded for large $\sigma\mathcal{A}$. To this effect, let us re-organize \eq{boskkp} in a suitable way. 
Using the asymptotic expansion of Bessel functions for large arguments:
\begin{equation}
\label{bfexp}
K_{j}(z) \sim \sqrt{\frac{\pi}{2z}}\ee^{-z}
\left(1 + \sum_{r=1}^\infty a_r^{(j)} z^{-r}\right)~,
\end{equation}
where the coefficients $a_r^{(j)}$ are well known, we obtain
\begin{equation}
\label{Idexp1}
\mathcal{I}^{(d)} = \sqrt{2\pi} \left(\frac{\sigma}{2\pi}\right)^{\frac{d-2}{2}}\, V_T \,
\sum_{k,k'=0}^\infty  c_k c_{k'} 
\left(\frac{\cale}{u}\right)^{\frac{d-2}{2}} \ee^{-\sigma \mathcal{A}\cale} \left(1 + \sum_{r=1}^\infty a_r^{(\frac{d-1}{2})} (\sigma\mathcal{A}\cale)^{-r}\right)~.
\end{equation}
We can then Taylor expand the expression \eq{bos19} of $\cale$ and write it in the form
\begin{equation}
\label{Xexp}
\cale = 1 + \sum_{t=1}^\infty d_t\left(a,b;u\right)(\sigma\mathcal{A})^{-t}~,
\end{equation}
having introduced, for notational simplicity,
\begin{equation}
\label{ab}
a = k + k' - \frac{d-2}{12}~,
\hskip 0.4cm
b = k - k'~.
\end{equation}
In particular one has
\begin{equation}
\label{d1}
d_1 = 2\pi u a = 2\pi u\left(k + k' - \frac{d-2}{12}\right)~.
\end{equation}
For the sake of readability, we leave to Appendix \ref{app:a} most of the details of the present computation, such as explicit expressions of the various expansion coefficients, describing here
only the basic steps.
 
Plugging the expansion \eq{Xexp} into \eq{Idexp1} we obtain
\begin{equation}
\label{iexp1}
\mathcal{I}^{(d)} \propto
\sigma^{\frac{d-2}{2}}\,
\frac{\ee^{-\sigma\mathcal{A}}}{u^{\frac{d-2}{2}}}\sum_{k,k'=0}^\infty  c_k c_{k'} 
\ee^{-2\pi u \left(k + k' - \frac{d-2}{12}\right)}
\left\{1 + \sum_{s=1}^\infty \frac{g_s\left(a,b;u\right)}{(\sigma\mathcal{A})^{s}}\right\}
~,
\end{equation}
where the coefficients $g_s$ can be re-constructed 
starting from the coefficients appearing in eq.s (\ref{bfexp}) and (\ref{Xexp}); 
see Appendix \ref{app:a} for more details.

We can introduce $Q = \exp(-2\pi u)$ and re-write \eq{iexp1} as follows:
\begin{equation}
\label{iexp2}
\mathcal{I}^{(d)} \propto
\sigma^{\frac{d-2}{2}}\,
\frac{\ee^{-\sigma\mathcal{A}}}{u^{\frac{d-2}{2}}} 
\lim_{\overline Q\to Q}
\sum_{k,k'=0}^\infty  c_k c_{k'}
\left\{1 + \sum_{s=1}^\infty \frac{g_s(a,b;u)}{(\sigma\mathcal{A})^{s}}\right\}
Q^{k - \frac{d-2}{24}}\overline Q^{k' - \frac{d-2}{24}}~.
\end{equation}
At this point, we can effectively replace
\begin{equation}
\label{subder}
a = k+k'- \frac{d-2}{12} \longrightarrow Q\frac{d~}{dQ} +  \overline Q\frac{d~}{d\overline Q}~,
\hskip 0.8cm
b = k - k' \longrightarrow Q\frac{d~}{dQ} -  \overline Q\frac{d~}{d\overline Q}
\end{equation}
and take into account  the definition, \eq{bos14}, of the coefficients $c_{k}$ to obtain
the form
\begin{equation}
\label{iexp3}
\mathcal{I}^{(d)} \propto 
\sigma^{\frac{d-2}{2}} \,
\frac{\ee^{-\sigma\mathcal{A}}}{u^{\frac{d-2}{2}}} 
\lim_{\overline Q\to Q} \left\{1 + \sum_{s=1}^\infty 
\frac{g_s\left(Q\frac{d~}{dQ} - \overline Q\frac{d~}{d\overline Q},Q\frac{d~}{dQ} + \overline Q\frac{d~}{d\overline Q} ;u\right)}{(\sigma\mathcal{A})^{s}}\right\}
\left[\eta(Q)\eta(\overline Q)\right]^{2-d}~,
\end{equation}
with, in the end, $Q$ being set to $\exp(-2\pi u)$. Finally, we can rewrite the above expression as
\begin{equation}
\label{iexp4}
\mathcal{I}^{(d)} \propto
\sigma^ {\frac{d-2}{2}}\,
\frac{\ee^{-\sigma\mathcal{A}}}{\left[\sqrt{u}
\eta^2(\ii u)\right]^{d-2}} \left\{1 + \sum_{s=1}^\infty \frac{f_s(u)}{(\sigma\mathcal{A})^{s}} \right\}
\end{equation}
where the coefficients $f_s$ are given by
\begin{equation}
\label{fcoeff}
f_s(u) = \eta^{2d-4}(\ii u) \lim_{\overline Q\to Q= \ee^{-2\pi u}} 
g_s\left(Q\frac{d~}{dQ} - \overline Q\frac{d~}{d\overline Q},Q\frac{d~}{dQ} + \overline Q\frac{d~}{d\overline Q} ;u\right)
\left[\eta(Q)\eta(\overline Q)\right]^{2-d}
\end{equation}
and can be related to Eisenstein series. In Appendix \ref{app:a} we derive the expression of the 2-nd loop and 3-rd loop 
coefficients, $f_1$ and $f_2$, but it would be straightforward to push the computation to higher orders.
Our expression for $f_1$ agrees completely with the (corrected) Dietz-Filk result \eq{fuis}. Let us remark 
that our result eq.(\ref{iexp4}) is proportional (with no need of ad hoc modifications) to $\sigma^ {\frac{d-2}{2}}$ , i.e. to $\sqrt{\sigma}$ in three dimensions. This 
result agrees with the (completely independent) field theoretical calculations in the framework of the $\phi^4$ model of \cite{m90,pv95}  and also, as we shall see, with the Monte Carlo simulations.

\section{Comparison with Monte Carlo data}
\label{sec:mc}
In a very recent publication \cite{nuovomc}, precise Monte Carlo data on the free energy $F_s$ of interfaces 
in the 3d Ising model were presented. The most accurate sets of data were obtained for square lattices
(i.e., in the case $u=1$), for values of the Ising coupling given by 
$\beta= 0.226102$ (Table 4 of \cite{nuovomc}; we shall refer to this data set as the set n. 1) and 
$\beta=0.236025$ (Table 2 of \cite{nuovomc}, set 2). 
The data set 1 contains 21 points, obtained at lattice sizes $L_1=L_2\equiv L$ 
ranging from $L=10$ to $L=30$; the 9 points in data set 2 are%
\footnote{We consider here only the data in Table 2 of \cite{nuovomc} which, for a given value of $L$, are 
obtained with $L_0$, the size of the lattice in the transverse direction, being equal to $3L$, to ensure
 uniformity with the data set 1, which is obtained with this choice.}
for $L=6$ to $L=14$. 

Previous work regarding other observables \cite{chp05} 
has made it clear that, in appropriate regimes, the 3d Ising model can be successfully described by an 
effective string theory. For the two values of $\beta$ which we study here very precise estimates for the 
string tension exist~\cite{Caselle:2004jq} (see Table 1 of 
\cite{nuovomc}). For set 1 one has $\sigma = 0.0105241$; for set 2,  $\sigma=0.044023$. This 
entails that the points in data set 1 correspond to values of $\sqrt{\sigma\mathcal{A}}$ ranging from $1.02$ 
to $3.07$, while the points in data set 2 correspond to values from $1.26$ to $2.94$.

\begin{table}
\begin{center}
\begin{tabular}{cccccc}
\hline
\hline 
& & \multicolumn{2}{c}{$N=100$} & \multicolumn{2}{c}{$N=0$} \\
\cline{3-6}
$L_{\mathrm{min}}$ & $(\sqrt{\sigma\mathcal{A}})_{\mathrm{min}}$ & $\mathcal{N}$ & $\chi^2/(\mathrm{d.o.f})$ 
& $\mathcal{N}$ & $\chi^2/(\mathrm{d.o.f})$ 
\\ 
\hline
\multicolumn{6}{c}{Data set 1}
\\
\hline
19 & 1.949 & 0.91957(18) & 4.22 & 0.91413(18) & 1.60\\ 
20 & 2.051 & 0.91891(22) & 1.84 & 0.91377(22) & 0.88\\ 
21 & 2.154 & 0.91836(27) & 0.63 & 0.91344(27) & 0.47\\ 
22 & 2.257 & 0.91829(33) & 0.70 & 0.91354(33) & 0.50\\ 
23 & 2.359 & 0.91797(45) & 0.63 & 0.91339(45) & 0.53\\ 
24 & 2.462 & 0.91762(57) & 0.57 & 0.91316(57) & 0.55\\ 
25 & 2.565 & 0.91715(75) & 0.50 & 0.91279(75) & 0.55\\
\hline
\multicolumn{6}{c}{Data set 2}
\\
\hline 
9 & 1.888 & 0.91052(21) & 7.22 & 0.90466(21) & 2.22 \\
10 & 2.098 & 0.90924(33) & 2.71 & 0.90413(33) & 1.69\\
11 & 2.308 & 0.90820(51) & 1.12 & 0.90349(51) & 1.33\\
\hline 
\hline
\end{tabular}
\end{center}
\caption{\small The fit of the NG free energy \eq{Fc0} with normalization $\mathcal{N}$ to the two data set of Ref. \cite{nuovomc}, performed using only the points in Table 4 and 2 of \cite{nuovomc} corresponding 
to lattice sizes $L\geq L_{\mathrm{min}}$, i.e., those with $\sqrt{\sigma\mathcal{A}}\geq (\sqrt{\sigma\mathcal{A}})_{\mathrm{min}}$.
The reduced $\chi^2$ becomes of order unity for $(\sqrt{\sigma\mathcal{A}})_{\mathrm{min}}\gtrsim 2$. 
The fit is performed by truncating the sum over the oscillator levels at $N=100$ or at $N=0$, i.e. keeping 
only the 0-mode contributions (last two columns).}
\label{tab:fits}
\end{table}

Using the information above, it is possible to compare the MC values of the free energy $F_s$ in data set 1
and 2 to the free energy $F$ corresponding to our partition function
\eq{boskkp} (in $d=3$, and for $u=1$), where we factor out the transverse volume factor $V_T$ and allow for
an overall normalization $\ee^{-\mathcal{N}}$:
\begin{equation}
\label{Fc0}
F = -\log \left(\frac{\mathcal{I}^{(3)}}{V_T}\right) + \mathcal{N}~.
\end{equation}
The constant $\mathcal{N}$ will be the only free parameter to be fitted to the data.
For numerical evaluation of the free energy \eq{Fc0}, we
re-arrange the sums appearing in \eq{boskkp} using the property 
\begin{equation}
\label{bos20}
\sum_{k,k'=0}^\infty c_k c_{k'}\,f((k-k')^2,k+k') 
=
\sum_{m=0}^\infty \sum_{k=0}^m c_k c_{m-k} f\left((2k-m)^2,m\right)~,
\end{equation}
valid for any function $f$ of the specified arguments, and write \eq{boskkp} as
\begin{equation}
\label{bos21}
\mathcal{I}^{(d)} = 2
\left(\frac{\sigma}{2\pi}\right)^{\frac{d-2}{2}}\, V_T \, \sqrt{\sigma\mathcal{A}u}
\sum_{m=0}^\infty \sum_{k=0}^m c_k c_{m-k} 
\left(\frac{\cale}{u}\right)^{\frac{d-1}{2}}\, K_{\frac{d-1}{2}}\left(\sigma\mathcal{A}\cale\right)~,
\end{equation}
where $\cale$ is now $\cale((2k-m)^2,m)$, 
since we transformed the summation indices as in \eq{bos20}. The integer $m$ represents the total (left-  
plus right-moving) oscillator number; the contributions to the partition function from states of a given $m$
are suppressed by exponentials of the type $\exp(-2\pi u\, m)$, as it can easily be seen from \eq{bos21} 
upon expanding the argument of the Bessel function using eq.s (\ref{Xexp},\ref{d1}).
We can approximate the exact expression of $F$ by truncating the sum over $m$ at a certain $N$, i.e.,
neglecting the contributions of string states with total occupation number larger that $N$. 

\begin{figure}
\begin{center}

\begin{picture}(0,0)%
\includegraphics{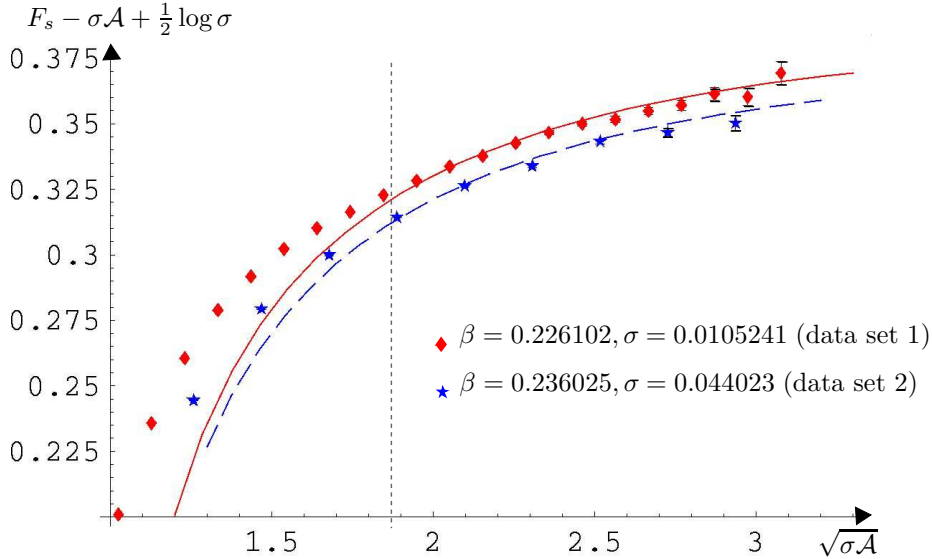}%
\end{picture}%
\setlength{\unitlength}{1184sp}%
\begin{picture}(18227,11603)(810,-10997)
\put(1217,222){\makebox(0,0)[lb]{\small $F_s - \sigma\mathcal{A} + \frac 12 \log \sigma$}}
\put(17776,-10711){\makebox(0,0)[lb]{\small $\sqrt{\sigma\mathcal{A}}$}}
\put(10276,-7391){\makebox(0,0)[lb]{\small $\beta=0.236025, \sigma= 0.044023$ (data set 2)}}
\put(10266,-6371){\makebox(0,0)[lb]{\small $\beta=0.226102, \sigma= 0.0105241$ (data set 1)}}
\end{picture}%

\end{center}
\caption{\label{fig:fit}
\small Two sets of Monte Carlo data for the interface free energy data provided in \cite{nuovomc} are compared to our theoretical predictions following from eq.s (\ref{Fc0},\ref{bos21}),
represented by the solid red line for the data set 1 and by the dashed blue line for data set 2. 
The only free parameter is the additive constant $\mathcal{N}$, corresponding to an overall normalization of 
the NG partition function, fitted to the data using the points to the right of the vertical grey dashed line
(see the text for more details). 
The error bars in the MC data are visible only for the rightmost points, but they were kept into
account in the fit. 
The sum over the level $m$ in \eq{bos21} was truncated at $N=100$.}
\end{figure}

Having fixed $u=1$ and having specified $N$, the free energy $F$ of \eq{Fc0} is just a function of $\sigma\mathcal{A}$, which we can fit to the MC points
in the two data sets referred to above.

The results of this comparison are summarized in Table \ref{tab:fits}. 
For each data set, we consider only the data corresponding to lattice sizes $L\geq L_{\mathrm{min}}$,
i.e., to values of $\sqrt{\sigma\mathcal{A}}\geq (\sqrt{\sigma\mathcal{A}})_{\mathrm{min}}$, 
and fix the value of the normalization constant $\mathcal{N}$ so as to minimize the $\chi^2$; we repeat
the analysis for various values of $L_{\mathrm{min}}$. Also, we consider the case where we
truncate the sum over the oscillator number at $N=100$ and the case $N=0$ where we keep only the zero-mode 
contributions. 

For both data sets 1 and 2, the reduced $\chi^2$ becomes of order one when
we consider in the fit only values of $\sqrt{\sigma\mathcal{A}}\gtrsim 2$. This is therefore the region 
in which our theoretical expression, derived from the Nambu-Goto model, describes well the data. 
This pattern is fully consistent with previous analysis of other observables in the 3d Ising models. In particular, the case of Polyakov loop correlators was recently discussed in
\cite{chp05}, where two different
behaviours were observed.  In the compact direction corresponding to the ``temperature", a good agreement with  the Nambu-Goto predictions was found
exactly in the same range of distances as in the present case. On the contrary,
in the direction with Dirichlet boundary conditions a clear disagreement 
was observed up to very large distances. This is probably due to some type of ``boundary effect" 
associated to the Dirichlet boundary conditions. It is exactly the absence of this type of effects which makes the case that we study in the present paper, periodic in both directions and hence free of boundary corrections, particularly well suited to study the effective string contributions, disentangled from other
spurious effects.

Considering the values of the $\chi^2$, we can take as our best estimates for the normalization constant 
$\mathcal{N}$ the value obtained at $L_{\mathrm{min}}=21$ for the data set 1, namely $\mathcal{N}=0.9184(3)$, 
and the value at $L_{\mathrm{min}}=11$ for data set 2, i.e. $\mathcal{N}=0.9082(5)$. It is
interesting to observe that truncating to $N=0$ the mode expansion gives
comparable values for the $\chi^2$ but leads to a sizeable change (with respect
to the statistical errors) of $\mathcal{N}$ which becomes
$\mathcal{N}=0.9134(3)$ and $\mathcal{N}=0.9035(5)$ for the data sets 1 and 2 respectively.
This difference is due to the fact that the truncation of the sum at $N=0$ replaces the area-independent 
1-loop determinant $\eta^{4-2d}(\ii u)$ appearing in the exact partition function \eq{iexp4} with 
its approximation $\exp(\pi u (d-2)/6)$. This can be seen by keeping the $m=0,k=0$ term only in \eq{bos21},
looking at the exponential behaviour $\exp(-\sigma\mathcal{A}\cale)$ and expanding $\cale$
to the first order, see Eq.s (\ref{Xexp}-\ref{d1}). In our situation, where $d=3$ and $u=1$, the 1-loop determinant (which is essentially reproduced summing the contributions up to $N=100$) contributes 
to the free energy a constant term $2 \log\left[\eta(\ii)\right] = - 0.527344...$ ; in the $N=0$ truncation, this constant becomes
$-\pi/6 = -0.523599$. The difference between these two values is $0.00374536...$, and  accounts for most of the difference between the estimates of the overall constant $\mathcal{N}$ at $N=100$ and $N=0$ given above.

It is instructive to compare our estimates for the overall constant with those obtained in \cite{nuovomc} by truncating the perturbative expansion in $1/\sigma \mathcal{A}$ to the second
order. Keeping into account the difference in the definition of the normalizations, which 
simply amounts to the contribution $2 \log\left[\eta^2(\ii)\right]$ of the one-loop determinant, we
would obtain, using the notations of~\cite{nuovomc},
$c_0+\frac12\log{\sigma}=0.3910(3)$ for the data set 1 and $N=100$, 
which should be compared with the value $c_0+\frac12\log{\sigma}=0.388(5)$. The estimates are compatible
within the errors. Since the fit we performed depends on one free parameter only, our estimate for
$c_0+\frac12\log{\sigma}$ turns out to be more precise than the one of
\cite{nuovomc}. As discussed in~\cite{nuovomc} the result we obtain is of the
order of magnitude but somehow larger than the prediction obtained in
~\cite{hp97,m90} in the framework of the $\phi^4$ theory:
$c_0+\frac12\log{\sigma}\sim0.29$.

To make the above discussion visually appreciable, in Figure \ref{fig:fit} we plot our theoretical curve 
against the MC points, having fixed the value of the 
normalization $\mathcal{N}$ according to the fit at $L_{\mathrm{min}}=21$ for the data set 1 and at 
$L_{\mathrm{min}}=11$ for data set 2 and having truncated the sum over the oscillators at $N=100$ in both cases. 
In fact, to draw a readable plot, it is convenient to subtract from both the Monte Carlo data and the 
theoretical prediction for the free energy the dominant scaling term $\sigma\mathcal{A}$
and a constant%
\footnote{This term is related to the natural appearance of a multiplicative factor
$\sigma^{\frac{d-2}{2}}$ in the effective string partition function, see \eq{bos21}. 
Furthermore we prefer to exhibit the same data plots appearing in \cite{nuovomc}.} 
term $-\frac 12 \log\sigma$. 

In Table 9 of ref. \cite{nuovomc} some data regarding rectangular lattices, i.e., 
with $u>1$ were also presented. Such data represent an important test of the form of our expression \ref{Fc0}. We present the 
results of this test in Table \ref{tab:rect}; they are quite encouraging, since we basically get 
agreement with the data within the
error bars, at least when including the contributions of higher oscillator numbers (up to $N=100$). 
Notice that in this test we use the value of the normalization constant $\mathcal{N}$ already determined 
by fitting it to the data for squared lattices of data set 2, see Table \ref{tab:fits}, since the data
for asymmetric lattices in ref. \cite{nuovomc} were obtained at exactly the same value of the coupling constant $\beta$ used for the data set 2. This makes the observed agreement even more remarkable, since it involved  no adjustable free parameter. 

It would clearly be desirable to obtain more extended sets of data for asymmetric lattices, to check
their degree of consistency with our Nambu-Goto predictions. 

\begin{table}[tb]
\begin{center}
\begin{tabular}{cccclll}
\hline
\hline 
$L_1$ & $L_2$ & $\sqrt{\sigma\mathcal{A}}$ & $u$ &  \multicolumn{1}{c}{$F_s$} & \multicolumn{1}{c}{diff $(N=100)$} 
& \multicolumn{1}{c}{diff $(N=0)$} \\
\hline 
$10$ & $12$ & $2.29843$ & $6/5$  & $\phantom{1}7.1670(6) $ &  $\phantom{-}0.0016$ & $0.0049$\\ 
$10$ & $15$ & $2.56972$ & $3/2$  & $\phantom{1}8.4449(12)$ &  $          -0.0004$ & $0.0042$\\ 
$10$ & $18$ & $2.81498$ & $9/5$  & $\phantom{1}9.6976(17)$ &  $          -0.0009$ & $0.0037$\\ 
$10$ & $20$ & $2.96725$ & $2$    & $          10.5235(25)$ &  $          -0.0012$ & $0.0035$\\ 
$10$ & $22$ & $3.11208$ & $11/5$ & $          11.3466(36)$ &  $\phantom{-}0.0017$ & $0.0064$\\ 
\hline
\end{tabular}
\end{center}
\caption{\small Comparison between the data presented in \cite{nuovomc} for rectangular lattices and our 
predictions.
The data were extracted at $\beta=0.236025$, corresponding to $\sigma=0.044023$, i.e., in the same situation as for the square lattices considered in data set 2 considered above. 
In the last columns we give the MC values  of the 
free energy $F_s$, according to Table 9 of \cite{nuovomc}, and the difference between these values and the 
values obtained from our expression \eq{Fc0}, with the sum over the oscillator number truncated at $N=100$ 
or at $N=0$. 
The normalization constant $\mathcal{N}$ was already determined by
the fit to the data set 2 presented in Table \ref{tab:fits}; we chose here the value obtained setting 
$L_{\mathrm{min}}=11$.}
\label{tab:rect}
\end{table}

One remarkable feature of the comparisons presented here in Tables \ref{tab:fits} and \ref{tab:rect} is 
that keeping just the zero modes of the string, i.e. setting $N=0$,
seems to yield a very good agreement with the data, in particular for symmetric interfaces. 
The contribution of non-zero modes modifies the overall normalization coming from the one-loop determinant, 
as discussed above, but it is much less important for higher loops, so that the shape of the free energy 
as a function of $\sigma\mathcal{A}$ is little changed. 

Keeping only the zero-modes, the quantity $\cale$ of \eq{bos19} which appears in the partition function
reduces effectively to
\begin{equation}
\label{Xzm}
\widehat \cale = \sqrt{1 - \frac{\pi u (d-2)}{3\sigma\mathcal{A}}}
\end{equation}
and the partition function \eq{bos21} simply becomes 
\begin{equation}
\label{boszero}
\widehat{\mathcal{I}}^{(d)} = 2
\left(\frac{\sigma}{2\pi}\right)^{\frac{d-2}{2}}\, V_T \, \sqrt{\sigma\mathcal{A}u} 
\left(\frac{\widehat \cale}{u}\right)^{\frac{d-1}{2}}\, K_{\frac{d-1}{2}}\left(\sigma\mathcal{A} \widehat \cale \right)~.
\end{equation}

\section{Conclusions}
\label{sec:conclusions}
In this paper, we studied the free energy of interfaces in a compact $d$-dimensional target space $T^d$, modeling them with a Nambu-Goto string effective action; this basically corresponds, for the $d=3$ case,
to use the capillary wave model. Performing standard covariant quantization and explicitly integrating over the modular parameter of the world-sheet, we were able to derive an exact expression for the sector of the NG partition function that describes the interface fluctuations .

Focusing on the three-dimensional case, we compared the exact predictions of the NG model with the free energy of interfaces obtained in \cite{nuovomc} through very precise Monte Carlo simulations in the Ising model. 
The Monte Carlo data are very well accounted for when we consider lattices of sufficiently large sizes, 
typically with $L\geq 2/\sqrt{\sigma}$; here $\sigma$ is the string tension, which can be associated with great accuracy to the value of the Ising coupling \cite{Caselle:2004jq}. 

Our analysis, which allows to compare the data with the exact theoretical prediction,
confirms and strengthens the evidence that the NG model is a very reliable effective model for such 
lattice sizes. Previously,
comparisons were made to the two-loop expansion of the NG model \cite{Dietz:1982uc}, which already gives a very good approximation for square interfaces; as pointed out in \cite{nuovomc}, however, and as confirmed here, higher order corrections are more important for asymmetric lattices. In fact, we found
extremely good agreement of the exact expression (which re-sums all the loop corrections) for the few data with $u\not=1$ presented in \cite{nuovomc}; 
it would be very interesting to obtain extended set of high precision Monte Carlo data for asymmetric lattices to check whether this agreement is confirmed, and to what extent.  

For smaller lattice sizes, rather large deviations from the NG predictions are observed in the data. This
pattern is in perfect agreement with what has been observed in previous studies of the NG predictions for other observables, such as the Polyakov loop correlator \cite{chp05}, and is in fact to be expected for deep
theoretical motivations. 

Indeed, we have studied the Nambu-Goto bosonic string in $d$ dimension 
paying no attention to the fact that 
in a consistent quantum treatment, when $d\not= 26$ further degrees of freedom have to be taken into account, beside the $d-2$ transverse oscillations. In the Polyakov first order formulation \cite{Polyakov:1981rd}, 
the scale $\ee^\phi$ of the world-sheet metric $h$ does not decouple, and gets in fact a Liouville-type action; in the Virasoro treatment, longitudinal modes of the string enter the game \cite{Goddard:1972iy,Brower:1972wj} (the two effects are  
not disconnected, see \cite{Marnelius:1986wp}).
For this reason these quantum bosonic string models have not been considered in the past 
\cite{Polchinski:1992vg,Polchinski:1991ax} 
as viable dual descriptions of situations, such as the confining regime of gauge theories or the fluid interfaces, where only transverse fluctuations are expected on physical grounds.

In a modern perspective, the extra degree of freedom corresponding to the Liouville mode could play somehow the r\^ole of the renormalization scale \cite{Polyakov:1997tj,Polyakov:1998ju}, in the spirit of (a non-conformal version of) AdS/CFT duality \cite{Maldacena:1997re}. 
It would be very interesting to try to take into account the contributions to the partition function
of the extra d.o.f. arising in $d=26$ at the quantum level, and to see whether they can account for the deviations of data, at small lattice sizes, from the prediction obtained via the na\"ive treatment of the NG model. Some effort in this direction, focusing on the relevance of the longitudinal modes for the description of large $N$ gauge theories, has been done in \cite{Dalley:2005fw}.

In \cite{Polchinski:1991ax} Polchinski and Strominger put forward a different proposal to circumvent the 
mismatch of d.o.f. between the bosonic string model and the transverse oscillations. An effective string was built, with an action analogous to the Polyakov one, where the independent scale 
$\ee^\phi$ of the metric gets replaced by the induced metric. This model is certainly very interesting, and it is receiving nowadays a renewed attention \cite{Drummond:2004yp,Kuti:2005xg}. It is noticeable that
the excitation spectra obtained within this model agree with the na\"ive NG ones up to two loops in the 
$1/(\sigma\mathcal{A})$ expansion; the prediction of this model would therefore nicely agree with NG in the region where the latter is valid. Higher corrections, whose computation does not appear to be simple,
could in principle better explain the data for smaller lattices.

In conclusion, there is a renewed interest in the problem of identifying the correct QCD string dual and of testing the various proposals, see for instance
\cite{Juge:2004xr,Drummond:2004yp,Brower:2005pb,Brower:2005cd,Kuti:2005xg,Dalley:2005fw}. As already said, 
our contribution corroborates
the ``experimental'' result that the na\"ive treatment of the NG model provides a very accurate description of fluctuating surfaces for sufficiently large sizes, and that clear deviations appear for smaller lattices which should be accounted for
by the correct dual model (at least down to some lower scale where the string description itself breaks down). 
On the theoretical side, our derivation of the exact partition function, and not just of its perturbative expansion, could offer some insight for a similar treatment of consistent models including also the extra d.o.f. arising in $d\not=26$, or of the Polchinski-Strominger string.

\acknowledgments{We thank F. Gliozzi, M. Hasenbusch, M. Panero and I. Pesando for useful discussions and comments.} 

\appendix
\section{Loop expansion of the exact result}
\label{app:a}
As discussed in section \ref{sec:bos_string} eq.(\ref{boskkp}),
the exact expression of the interface partition function
\begin{equation}
\label{eq:I(d)}
\mathcal{I}^{(d)}=2\left(\frac{\sigma}{2\pi}\right)^{\frac{d-2}{2}}V_{T}\sqrt{\sigma
\mathcal{A} u}
\sum_{m=0}^{\infty}\sum_{k=0}^{m}c_{k}c_{m-k}\left(\frac{\cale}{u}\right)^{\frac{d-1}{2}}K_{\frac{d-1}{2}}(\sigma
\mathcal{A} \cale)
\end{equation}
has to be expanded for large $\sigma\mathcal{A}$, in order to compare it with the result of the functional integral approach. Therefore we have to use the asymptotic expression eq. (\ref{bfexp}) of the modified Bessel functions  for large argument which reads, making explicit the first coefficients
$a_{1}^{(j)}$ and $a_{2}^{(j)}$, 
\begin{equation*}
K_{j}(z)\sim\sqrt{\frac{\pi}{2z}}e^{-z}\left\{
1+\frac{4j^{2}-1}{8z}+\frac{(4j^{2}-1)(4j^{2}-9)}{2!(8z)^{2}}+...\right\}~.
\end{equation*}
Inserting the expansion of Bessel functions in eq. (\ref{eq:I(d)}) gives \eq{Idexp1} (which we repeat here for commodity):
\begin{equation}
\label{expansionI}
\mathcal{I}^{(d)}=\sqrt{2\pi}\left(\frac{\sigma}{2\pi}\right)^{\frac{d-2}{2}}V_{T}\sum_{k,k'=0}^{\infty}c_{k}c_{k'}
\left(\frac{\cale}{u}\right)^{\frac{d-2}{2}}e^{-\sigma
\mathcal{A}\cale}\left(1+\sum_{r=1}^{\infty}a_{r}^{(\frac{d-1}{2})}(\sigma\mathcal{A}\cale)^{-r}\right)~.
\end{equation}
The expression of $\cale$ was given in \eq{bos19}, and can be rewritten as
\begin{equation*}
\cale(a,b)=\sqrt{1 + \frac{4\pi\, u}{\sigma \mathcal{A}}a +
\frac{4\pi^{2} \, u^2\,}{(\sigma\mathcal{A})^2}b^{2}}
\end{equation*}
in terms of $a$ and $b$ defined as in \eq{ab}:
\begin{equation}
a = k + k' - \frac{d-2}{12}~, \hskip 0.4cm b = k - k'~.
\end{equation}
We expand $\cale$ as:
\begin{equation*}
\cale(a,b)\sim 1 + \frac{u}{\sigma \mathcal{A}}2\pi a +
\frac{u^{2}}{(\sigma\mathcal{A})^{2}}2\pi^{2}\left(b^{2}-a^{2}\right)+\frac{u^{3}}{(\sigma\mathcal{A})^{3}}4\pi^{3}a\left(a^{2}-b^{2}\right)+...
\end{equation*}
and insert this expansion into eq. (\ref{expansionI}), obtaining
\begin{eqnarray*}
\mathcal{I}^{(d)}&\propto& \sigma^{\frac{d-2}{2}}\frac{e^{-\sigma
\mathcal{A}}}{u^{\frac{d-2}{2}}}\sum_{k,k'=0}^{\infty} c_{k}
c_{k'} e^{-2\pi u \left(k+k'-\frac{d-2}{12}\right)}
 \\
&\times&  \left\{1+\frac{u}{\sigma \mathcal{A}}(d-2)\pi
a+\frac{u^2}{(\sigma \mathcal{A})^{2}}
(d-2)\left[(d-6)\frac{\pi^{2}}{2}a^{2}+\pi^{2}b^{2}\right]+\mathcal{O}\left(\frac{1}{(\sigma\mathcal{A})^{4}}\right)\right\}
\\
&\times&  \left\{1+\frac{u^{2}}{\sigma\mathcal{A}}2\pi^{2}\left(a^{2}-b^{2}\right)
- \frac{4\pi^{3} u^3 a\left(a^{2}-b^{2}\right) -
2\pi^{4} u^4 \left(b^{2}-a^{2}\right)^{2} }{(\sigma\mathcal{A})^{2}} +
\mathcal{O}\left(\frac{1}{(\sigma\mathcal{A})^{3}}\right)\right\} \\
&\times& \left\{ 1+\frac{1}{\sigma \mathcal{A}}\frac{d(d-2)}{8} -
\frac{\frac{d(d-2)(d^{2}-2d-8)}{128} - \frac{\pi u\,d(d-2)a}{4}}{(\sigma\mathcal{A})^{2}} 
+\mathcal{O}\left(\frac{1}{(\sigma\mathcal{A})^{3}}\right)\right\}.
\end{eqnarray*}

The result up to the third order is thus of the form of \eq{iexp1}: 
\begin{equation}
\label{expansion} \mathcal{I}^{(d)}\propto
\sigma^{\frac{d-2}{2}}\frac{e^{-\sigma
\mathcal{A}}}{u^{\frac{d-2}{2}}}\sum_{k,k'=0}^{\infty} c_{k}
c_{k'} e^{-2\pi u
\left(k+k'-\frac{d-2}{12}\right)}\left\{1+\frac{g_{1}}{\sigma\mathcal{A}}+\frac{g_{2}}{(\sigma\mathcal{A})^{2}}\right\},
\end{equation}
with $g_{1}$ and $g_{2}$ explicitly given by
\begin{eqnarray}
\label{g1}
g_{1}&=& \frac{d(d-2)}{8}+
u(d-2)\pi a-u^{2}2\pi^{2}\left(b^{2}-a^{2}\right)~,\\
\label{g2}
g_{2}&=& \left[\frac{d(d-2)(d^{2}-2d-8)}{128}
\right] +
u\frac{d(d-2)(d-4)\pi}{8}a+u^{2}\frac{(d-2)(d-4)\pi^{2}}{4}\left(3a^{2}-b^{2}\right)\nonumber\\
&+& u^{3} 2\pi^{3}(d-4)a
\left(a^{2}-b^{2}\right)+u^{4}2\pi^{4}\left(b^{2}-a^{2}\right)^{2}.
\end{eqnarray}
Introducing the new variable $Q=\exp(-2\pi u)$ we can write eq.
(\ref{expansion}) in the form
\begin{equation}
\mathcal{I}^{(d)}\propto \sigma^{\frac{d-2}{2}}\frac{e^{-\sigma
\mathcal{A}}}{u^{\frac{d-2}{2}}}\lim_{\overline{Q}\rightarrow
Q}\left\{1+\frac{g_{1}}{\sigma\mathcal{A}}+\frac{g_{2}}{(\sigma\mathcal{A})^{2}}\right\}[\eta(Q)\eta(\overline{Q})]^{2-d}~,
\end{equation}
where we have replaced $a$ and $b$ with the derivatives on $Q$ and $\overline{Q}$ (see eq. (\ref{subder})):
\begin{eqnarray}
\label{ainQ}
a&\rightarrow& Q\frac{d}{dQ}+\overline{Q}\frac{d}{d\overline{Q}}~,\\
b&\rightarrow& Q\frac{d}{dQ}-\overline{Q}\frac{d}{d\overline{Q}}~.
\end{eqnarray}
As we can see from the expansion coefficients (\ref{g1}) and (\ref{g2}), the only terms involving powers of $a$ and $b$ higher than 1 are
\begin{eqnarray*}
b^{2}-a^{2}&=&-4Q\frac{d}{dQ}\overline{Q}\frac{d}{d\overline{Q}}\\
3a^{2}-b^{2}&=& 2\left(Q\frac{d}{dQ}\right)^{2}+2\left(\overline{Q}\frac{d}{d\overline{Q}}\right)^{2}+8Q\frac{d}{dQ}\overline{Q}\frac{d}{d\overline{Q}}\\
\left(b^{2}-a^{2}\right)^{2}&=& 16\left(Q\frac{d}{dQ}\right)^{2}\left(\overline{Q}\frac{d}{d\overline{Q}}\right)^{2}
\end{eqnarray*}
which have to be applied to $\eta^{2-d}(Q)\eta^{2-d}(\overline{Q})$.
Applying the expressions \eq{der1eta} and \eq{der2eta} of the first logarithmic derivatives of the eta-function
given in Appendix \ref{app:useful}, we have
\begin{eqnarray*}
Q\frac{d}{dQ}\eta^{2-d}(Q)&=&(2-d) \eta^{2-d}(Q)
\frac{E_{2}(Q)}{24}~,\\
\left(Q\frac{d}{dQ}\right)^{2}\eta^{2-d}(Q)&=&(2-d)\eta^{2-d}(Q)\frac{(4-d)E_{2}^{2}(Q)-2E_{4}(Q)}{576}~.
\end{eqnarray*} 
It is easy to check that, after the application of the derivatives, a factor of $\eta^{4-2d}(\ii u)$ 
will always appear in front of the various terms and
the total result up to the third order can be written in the following form:
\begin{equation*}
\mathcal{I}^{(d)}\propto\sigma^{\frac{d-2}{2}}\frac{\ee^{-\sigma
\mathcal{A}}}{u^{\frac{d-2}{2}}}\frac{1}{\eta^{2d-4}(\ii u)}
\left\{1+\frac{f_{1}}{\sigma\mathcal{A}}+\frac{f_{2}}{(\sigma\mathcal{A})^{2}}\right\}~.
\end{equation*}
First of all we consider the $f_{1}$ term derived from the $g_{1}$ one. As one can see, the only derivative involved in the computation is $Q\frac{d}{dQ}$: this implies that our final formula will include $E_{2}$ functions but no $E_{4}$. We can divide the calculation in
function of the powers of $u$. The term independent from it is the first one appearing in eq. (\ref{g1}):
\begin{equation*}
f_{1,0}=\frac{d(d-2)}{8}.
\end{equation*}
The next power of $u$, after the substitution (\ref{ainQ}), gives, acting on the $\eta$ functions,
\begin{equation*}
f_{1,1}= - 2\pi u
(d-2)^{2}\frac{E_{2}(\ii u)}{24}.
\end{equation*}
The last term is instead:
\begin{equation*}
f_{1,2}= 2\pi^{2} u^2 (d-2)^{2}\frac{E_{2}^{2}(\ii u)}{144}.
\end{equation*}
The final result is:
\begin{equation}
\label{f1}
f_{1} =
\left\{\frac{(d-2)^{2}}{2}\left[\left(\frac{\pi}{6}\right)^{2}u^{2}E_{2}^{2}(\ii u)-
\frac{\pi}{6}uE_{2}(\ii u) \right]+\frac{d(d-2)}{8}\right\}.
\end{equation}
Proceeding in the same way to evaluate $f_{2}$ we find:
\begin{eqnarray}
f_{2}&=&
\left\{u^{4} \frac{\pi^{4}(d-2)^{2}}{18} \left[\frac{(4-d)E_{2}^{2}-2E_{4}}{24}\right]^{2}
+u^{3} \frac{\pi^{3}}{72}(d-4)(d-2)^{2} \left[\frac{E_{2}^{3}}{12}(4-d)\right.\right. 
\nonumber\\
&-& \left.\left.\frac{E_{2}E_{4}}{6}\right]
+u^{2}\frac{\pi^{2}(d-2)^{2}(d-4)}{4}\left[\frac{E_{4}}{72}+\frac{(3d-8)E_{2}^{2}}{144}\right]
\right.
\nonumber\\
&-& \left. u\frac{\pi}{96}d(d-2)^{2}(d-4)E_{2}+\frac{1}{128}d(d-2)(d^{2}-2d-8) \right\}
\end{eqnarray}
(all modular forms above have to be evaluated at $\ii u$).

\section{The two loop contribution: functional integral computation}
\label{app:2loop}
In this appendix we check the calculation of the two-loop terms of
the free energy made by Dietz and Filk with the functional
integral method. Starting from the partition function written in
the physical gauge and expanding it in powers of $\frac{1}{\sigma
\mathcal{A}}$, one can evaluate it on a rectangular domain $B$ of sizes $(R,T)$.
Defining $\mathbf{G}$ as the
inverse Laplacian of the theory, the second order terms are given, in the notation of \cite{Dietz:1982uc},
by:
\begin{eqnarray}
\label{eq:Z_{2}}
{Z}_{\Gamma,B}^{(2)} &=& \Bigl[\lambda_{1} D (D-1)(\langle1\rangle-\langle2\rangle)+\lambda_{2} D (\langle3\rangle+\langle4\rangle+ 2 \langle1\rangle - 4 \langle2\rangle) \nonumber \\
&+& \frac{\lambda_{2}}{3} D (D-1)
(\langle3\rangle+\langle4\rangle+ 6 \langle1\rangle)\Bigr]~,
\end{eqnarray}
where $D$ is $d-2$ in our notation, $\Gamma$ speficies the topology of the boundary $B$; we are 
interested in the case where the topology is $S_1\times S_1$. 
$\lambda_{1}$ and $\lambda_{2}$ are parameters which depend on the string model under consideration 
and in particular they are given by
\begin{equation}
\lambda_{1}=1,
\hskip 0.4cm
\lambda_{2}=-\frac{1}{4}
\end{equation}
in the Nambu-Goto case. Moreover, the following definitions are used:
\begin{eqnarray*}
\langle1\rangle &=& \frac{1}{\sigma}\int_{0}^{R} dz \int_{0}^{T}
dt \frac{\partial^{2}\mathbf{G}}{\partial z \partial
z'}|_{z=z',t=t'} \frac{\partial^{2}\mathbf{G}}{\partial t \partial
t'}|_{z=z',t=t'}~,\\
\langle2\rangle &=& \frac{1}{\sigma}\int_{0}^{R} dz \int_{0}^{T}
dt \left(\frac{\partial^{2}\mathbf{G}}{\partial z
\partial
t'}\right)^{2}|_{z=z',t=t'}~, \\
\langle3\rangle &=& \frac{3}{\sigma}\int_{0}^{R} dz \int_{0}^{T}
dt \left(\frac{\partial^{2}\mathbf{G}}{\partial z \partial
z'}\right)^{2}|_{z=z',t=t'}~, \\
\langle4\rangle &=& \frac{3}{\sigma}\int_{0}^{R} dz \int_{0}^{T}
dt \left(\frac{\partial^{2}\mathbf{G}}{\partial t \partial
t'}\right)^{2}|_{z=z',t=t'}~.
\end{eqnarray*}
Notice that the formul\ae used by Dietz and Filk in eq. (3.4) have to be multiplied by $-\frac{1}{2}$ to get 
the right normalization of their final result eq. (3.7). 

We shall deal with the divergent terms within the $\zeta$-function regularization scheme, following Dietz and Filk.
In the Nambu-Goto model with  $S_{1}\times S_{1}$ boundary
conditions on the rectangular domain, the Green function is:
\begin{equation*}
\mathbf{G} = \frac{1}{4\pi^{2}RT} \sum_{\atopnew{m,n=-\infty}{(m,n)\neq (0,0)}}^{+\infty} \frac{\exp{\left\{2\pi \ii
\frac{m}{T}(t-t')\right\}}\exp{\left\{2\pi \ii
\frac{n}{R}(z-z')\right\}}}
{\frac{n^2}{R^{2}}+\frac{m^{2}}{T^{2}}}~.
\end{equation*}
Let us now evaluate the various terms of
${Z}_{S_1\times S_1,(R,T)}^{(2)}$. To begin with, we find
\begin{eqnarray}
\label{1prov}
\langle1\rangle &=& \frac{1}{\sigma R T}  \sum_{\atopnew{m,n,p,q=-\infty}{\atopnew{(m,n)\neq (0,0)}{(p,q)\neq (0,0)}}}^{+\infty}  \frac{\frac{n^{2}}{R^{2}}\frac{q^{2}}{T^{2}}} {\left( \frac{n^{2}}{R^{2}}+\frac{m^{2}}{T^{2}}\right)\left( \frac{p^{2}}{R^{2}}+\frac{q^{2}}{T^{2}}\right)} \nonumber\\
&=&\frac{1}{\sigma R T} \left[ 16 \sum_{m,n,p,q=1}^{+\infty}
\frac{\frac{n^{2}}{R^{2}}\frac{q^{2}}{T^{2}}} {\left(
\frac{n^{2}}{R^{2}}+\frac{m^{2}}{T^{2}}\right)\left(
\frac{p^{2}}{R^{2}}+\frac{q^{2}}{T^{2}}\right)} + (m=0,n \neq
0,p,q) \right. \nonumber\\
&+& \left.  (m,n,p=0,q \neq 0)-(m=0,n\neq 0,p=0,q\neq 0) \right]~,
\end{eqnarray}
where with $(m=0,n \neq 0,p,q)$ we indicate the sum over any $n$
different from zero, any $p$ and any $q$ when $m$ has been fixed
at zero. The last term in r.h.s. of eq. (\ref{1prov}) is to avoid
the double counting and the first one can be rewritten using the
equality
\begin{equation}
\label{equality}
\sum_{m,n=1}^{\infty}\frac{\frac{n^{2}}{R^{2}}}{\frac{n^{2}}{R^{2}}+\frac{m^{2}}{T^{2}}}=E_{2}\left(
\ii\frac{R}{T}\right) \frac{\pi}{24} \frac{R}{T}~.
\end{equation}
The result is
\begin{eqnarray*}
\langle1\rangle &=& \frac{1}{\sigma R T} \left[ \left(
\frac{\pi}{6}\right)^{2} E_{2}\left(\ii\frac{R}{T}\right)
E_{2}\left(\ii\frac{T}{R}\right) +(m=0,n \neq 0,p,q)  \right. \\
&+& \left. (m,n,p=0,q\neq 0) - (m=0,n \neq 0,p=0, q\neq 0)\right]~.
\end{eqnarray*}

The last three terms can be evaluated using (\ref{equality}) and the $\zeta$ function regularization and they sum up to
\begin{equation*}
-\frac{\pi}{6}\frac{R}{T}E_{2}\left(\ii\frac{R}{T}\right)
-\frac{\pi}{6}\frac{T}{R}E_{2}\left(\ii\frac{T}{R}\right)+1~,
\end{equation*}
which vanishes using the modular property (\ref{eq:modular E_{2}}) for the
Eisenstein series $E_{2}$.
We get thus
\begin{equation*}
\langle1\rangle=\frac{1}{\sigma R T} H~,
\end{equation*}
where we have defined for notational convenience
\begin{equation}
\label{H}
H=\left( \frac{\pi}{6}\right)^{2} E_{2}\left(
\ii\frac{T}{R}\right) E_{2}\left(\ii\frac{R}{T}\right)~.
\end{equation}

The  contribution  $\langle2\rangle$ vanishes because the terms in the sum are odd:
\begin{equation*}
\langle2\rangle=\frac{1}{\sigma RT}
\sum_{\atopnew{m,n,p,q=-\infty}{\atopnew{(m,n)\neq(0,0)}{(p,q)\neq(0,0)}}}^{+\infty}\frac{mnpq}{\left(\frac{n^{2}}{R^{2}}+\frac{m^{2}}{T^{2}}\right)\left(\frac{p^{2}}{R^{2}}+\frac{q^{2}}{T^{2}}\right)}=0~.
\end{equation*}

To compute the expression $(\langle3\rangle+\langle4\rangle+6\langle1\rangle)$ which appears in (\ref{eq:Z_{2}}), one can rewrite it as a perfect square:
\begin{eqnarray*}
(\langle3\rangle+\langle4\rangle+6\langle1\rangle)&=&\frac{3}{\sigma}\int_{0}^{R}dz\int_{0}^{T}dt\left(\frac{\partial^{2}\mathbf G}{\partial z\partial z'}+\frac{\partial^{2}\mathbf G}{\partial t \partial t'}\right)^{2}|_{z=z',t=t'}\\
&=&\frac{3}{\sigma RT}\left[\sum_{\atopnew{m,n=-\infty}{(m,n)\neq(0,
0)}}^{+\infty}\frac{\frac{n^{2}}{R^{2}}+\frac{m^{2}}{T^{2}}}{\frac{n^{2}}{R^{2}}+\frac{m^{2}}{T^{2}}}\right]^{2}
= \frac{3}{\sigma RT}~.
\end{eqnarray*}

Collecting the previous results we obtain the following system of
equations:
\begin{eqnarray*}
\langle1\rangle &=& \frac{H}{\sigma RT}~, \\
\langle2\rangle &=& 0~, \\
\langle3\rangle + \langle4\rangle + 6\langle1\rangle &=& \frac{3}{\sigma RT}
\end{eqnarray*}
which gives, for the last term in equation (\ref{eq:Z_{2}}):
\begin{equation*}
\langle3\rangle+\langle4\rangle+2\langle1\rangle -4\langle2\rangle
= \frac{3-4H}{\sigma RT}~.
\end{equation*}

The total second order correction is therefore
\begin{eqnarray}
{Z}_{S_1\times S_1,B} &=& \frac{1}{\sigma R T} \left\{ D (D-1) H-\frac{1}{4} D (3-4H) -\frac{3}{12} D (D-1)\right\} \nonumber\\
&=& \frac{1}{\sigma R T} \left\{ D^{2} H -\frac{1}{4} D
(D+2)\right\}~.
\end{eqnarray}
This corresponds to our result%
\footnote{Dietz and Filk found: $ {Z}_{S_1\times S_1,(R,T)}^{(2)}
= \frac{1}{\sigma R T} \left\{D^{2} H -\frac{1}{4} [ D
(4D-1)]\right\}$} eq. (\ref{f1}) where $D=d-2$, up to an overall factor of $-\frac{1}{2}$.

\section{Useful formul\ae}
\label{app:useful}
In this appendix we collect some useful formul\ae.

\paragraph{Dedekind $\eta$ function}
The Dedekind eta function
is defined, in terms of the quantity $q=\exp\{2\pi \ii \tau\}$, by
\begin{equation} 
\label{defeta1}
\eta(\tau)=q^{\frac{1}{24}} \prod_{n=1}^{\infty}
\left( 1-q^{n} \right)~.
\end{equation}
One can expand it in $q$-series:
\begin{equation}
\label{defeta2}
[\eta(\tau)]^{-1}=\sum_{k=0}^{\infty} p_{k}\, q^{k-\frac{1}{24}}~,
\end{equation}
where $p_{k}$'s are the number of
partitions of $k$. In the text, we often switch between the notation $\eta(\tau)$ and $\eta(q)$ for the function defined in eq.s (\ref{defeta1},\ref{defeta2}), according to the convenience.

Under the modular transformations $T$ and $S$ the
Dedekind eta function transforms in the following way:
\begin{eqnarray}
\label{etaT}
\eta(\tau+1)&=& \ee^{\frac{\ii\pi}{12}} \eta(\tau)~, \\
\label{etaS}
\eta(-\frac{1}{\tau})&=& (-\ii\tau)^{\frac{1}{2}} \eta(\tau)~.
\end{eqnarray}
\paragraph{Eisenstein series}
The second Eisenstein function is defined by
\begin{equation*}
E_{2}(\tau)=1-24\sum_{n=1}^{\infty}\sigma_{1}(n)q^{n}=1-24\sum_{k=1}^{\infty}\frac{kq^{k}}{1-q^{k}}~,
\end{equation*}
where $\sigma_{1}(n)$ denotes the sum of the positive divisors of
$n$.
An useful property is
\begin{equation}
q\frac{d}{dq}E_{2}(\tau)=\frac{E_{2}^{2}(\tau)-E_{4}(\tau)}{12}~.
\end{equation}
where the fourth Eisenstein series is defined, in terms of the sum
of the cubes of the positive divisors of $n$, $\sigma_{3}(n)$, as
\begin{equation*}
E_{4}(\tau)=1+240\sum_{n=1}^{+\infty}\sigma_{3}\left(n\right)q^{n}~.
\end{equation*}
The modular properties of the Eisenstein functions are:
\begin{eqnarray}
\label{eq:modular E_{2}}
E_{2}(\tau) & = &\left(\frac{1}{\tau}\right)^{2}E_{2}\left(-\frac{1}{\tau}\right)+\frac{6}{\pi}
\frac{\ii}{\tau}~,
\\
E_{2k}(\tau)& = & (-1)^{k}\left(\frac{1}{\tau}\right)^{2k}E_{2k}\left(-\frac{1}{\tau}\right),k\geq 2~.
\end{eqnarray}
Multiple logarithmic derivatives of the eta function are related to Eisenstein series, and in particular we have:
\begin{eqnarray}
\label{der1eta}
q\frac{d}{dq}\eta^{\alpha}(\tau)&=&\alpha \eta^{\alpha}(\tau)
\frac{E_{2}(\tau)}{24}~,\\
\label{der2eta}
\left(q\frac{d}{dq}\right)^{2}\eta^{\alpha}(\tau)&=&\alpha\eta^{\alpha}\frac{(\alpha+2)E_{2}(\tau)^{2}-2E_{4}(\tau)}{576}~.
\end{eqnarray}
\paragraph{Poisson resummation formula}
In section \ref{sec:bos_string} the following resummation formula plays a key role:
\begin{equation}
\label{poisson_res}
\sum_{n=-\infty}^{+\infty}\exp\{-\pi
an^{2}+2\pi
ibn\}=a^{-\frac{1}{2}}\sum_{m=-\infty}^{+\infty}\exp\left\{
-\frac{\pi(m-b)^{2}}{a}\right\}
\end{equation}

\end{document}